\documentclass[lettersize,journal]{IEEEtran}
\usepackage{amsmath,amsfonts}
\usepackage{bm}
\usepackage{amssymb}
\usepackage{algorithmic}
\usepackage{algorithm}
\usepackage{multirow} 
\usepackage{makecell}
\usepackage{makecell,rotating,multirow,diagbox}
\usepackage{array}
\usepackage{textcomp}
\usepackage{stfloats}
\usepackage{url}
\usepackage{verbatim}
\usepackage{graphicx}
\usepackage{epstopdf}
\usepackage{subcaption}
\usepackage{cite}
\newtheorem{lemma}{Lemma}
\newtheorem{prop}{Proposition}
\begin{document}
\title{Multi-Functional RIS Integrated Sensing and Communications for 6G Networks}
\author{
	    Dongsheng Han,~\IEEEmembership{Member,~IEEE,}
        Peng Wang,
        Wanli Ni, 
        Wen Wang,
        Ailing Zheng,\\
        Dusit Niyato,~\IEEEmembership{Fellow,~IEEE},
        and Naofal Al-Dhahir,~\IEEEmembership{Fellow,~IEEE}     
  \thanks{      
        This research is supported in part by the S$\&$T Program of Hebei under Grant SZX2020034, and in part by the National Research Foundation, Singapore, and Infocomm Media Development Authority under its Future Communications Research $\&$ Development Programme, Defence Science Organisation (DSO) National Laboratories under the AI Singapore Programme (FCP-NTU-RG-2022-010 and FCP-ASTAR-TG-2022-003), Singapore Ministry of Education (MOE) Tier 1 (RG87/22), the NTU Centre for Computational Technologies in Finance (NTU-CCTF), and Seitee Pte Ltd.}
\thanks{
	Dongsheng Han and Peng Wang are with the School of Electrical and Electronic Engineering, North China Electric Power University, Beijing 102206, China, and also with Hebei Key Laboratory of Power Internet of Things Technology, North China Electric Power University, Baoding 071003 China (e-mail: handongsheng@ncepu.edu.cn, wangpeng9712@ncepu.edu.cn).}
\thanks{
	Wanli Ni is with the Department of Electronic Engineering, Tsinghua University, Beijing 100084, China (e-mail: niwanli@tsinghua.edu.cn).}
\thanks{
	Wen Wang is with the Pervasive Communications Center, Purple Mountain Laboratories, Nanjing 211111, China (email: wangwen@pmlabs.com.cn).}	
\thanks{
	Ailing Zheng is with State Key Laboratory of Networking and Switching Technology, Beijing University of Posts and Telecommunications, Beijing 100876, China (ailing.zheng@bupt.edu.cn).}
\thanks{
	Dusit Niyato is with the School of Computer Science and Engineering, Nanyang Technological University, Singapore 639798 (e-mail: dniyato@ntu.edu.sg).}
\thanks{
	Naofal Al-Dhahir is with the Department of Electrical and Computer Engineering, The University of Texas at Dallas, USA (email: aldhahir@utdallas.edu).}
}
\maketitle

\begin{abstract}
 In this paper, we propose a novel multi-functional reconfigurable intelligent surface (MF-RIS) that supports signal reflection, refraction, amplification, and target sensing simultaneously. Our MF-RIS aims to enhance integrated communication and sensing (ISAC) systems, particularly in multi-user and multi-target scenarios. Equipped with reflection and refraction components (i.e., amplifiers and phase shifters), MF-RIS is able to adjust the amplitude and phase shift of both communication and sensing signals on demand. Additionally, with the assistance of sensing elements, MF-RIS is capable of capturing the echo signals from multiple targets, thereby mitigating the signal attenuation typically associated with multi-hop links. We propose a MF-RIS-enabled multi-user and multi-target ISAC system, and formulate an optimization problem to maximize the signal-to-interference-plus-noise ratio (SINR) of sensing targets. This problem involves jointly optimizing the transmit beamforming and MF-RIS configurations, subject to constraints on the communication rate, total power budget, and MF-RIS coefficients. We decompose the formulated non-convex problem into three sub-problems, and then solve them via an efficient iterative algorithm. Simulation results demonstrate that: 1) The performance of MF-RIS varies under different operating protocols, and energy splitting (ES) exhibits the best performance in the considered MF-RIS-enabled multi-user multi-target ISAC system; 2) Under the same total power budget, the proposed MF-RIS with ES protocol attains $\rm{52.2}\%$, $\rm{73.5}\%$, and $\rm{60.86}\%$ sensing SINR gains over active RIS, passive RIS, and simultaneously transmitting and reflecting RIS (STAR-RIS), respectively; 3) The number of sensing elements will no longer improve sensing performance after exceeding a certain number.
\end{abstract}

\section{Introduction}
 \IEEEPARstart{W}{ith} the explosive growth of Internet of Things (IoT) devices, spectrum scarcity has become increasingly severe in the fields of communications and radar. At the same time, the emergence of numerous environment-aware applications, such as vehicle-to-everything (V2X), virtual reality (VR), and augmented reality (AR), putting forward higher requirements for both communication quality and sensing accuracy. To alleviate the exacerbated spectrum scarcity and provide ubiquitous wireless connectivity for emerging services, integrated sensing and communication (ISAC) has been proposed, which sparks significant interest from both academia and industry due to the dual functions of communication and sensing \cite{10158711, 10251151, 10038611}.  ISAC tightly integrates sensing and communication functions by sharing existing hardware facilities, spectrum, and signal processing frameworks, effectively improving spectrum efficiency while reducing hardware costs \cite{10180051, 10032141}. On the other hand, triggered by benefits from integration gain and coordination gain, ISAC can provide high-quality wireless communication and high-precision sensing. ISAC is expected to be a promising technique for the upcoming sixth-generation (6G) wireless systems.\par 
 However, ISAC still faces several challenges in practical scenarios, such as blocked line-of-sight (LoS) link, limited coverage, and significant pathloss \cite{9945983, 9965407}. Fortunately, the rise of reconfigurable intelligent surface (RIS) provides a novel approach to address these challenges \cite{9999292,123456}. With its ability to intelligently manipulate the wireless propagation environment, RIS can adjust the amplitude and phase shift of the incident signals on demand, thus creating efficient LoS links for sensing and communication \cite{789456,10254508,10336755}. The deployment of RIS achieves extended coverage and enhanced transmission quality by providing additional degrees of freedom (DoFs) and fading compensation, thereby improving both sensing and communication performance of ISAC systems\cite{10478980}.
\subsection{Related Work}
Recently, there have been some works introducing RIS into ISAC systems to improve sensing and communication performances for various ISAC scenarios \cite{9364358,10319318,9909807,10364735,10423585,10197455}. Specifically, the authors of \cite{9364358} considered a single-user and single-target ISAC scenario, where the RIS is deployed to tackle the issue of severe radar performance degradation caused by path loss. Furthermore, the authors of \cite{10319318} deployed an active RIS to assist the multi-user ISAC scenario. Compared to conventional passive RISs, active RIS can significantly mitigate harmful multiplicative fading over a multi-hop sensing link. 	Different from a conventional relay where the received signals need be decoded, processed, and regenerated, active RIS mainly manipulates the propagation direction of electromagnetic waves by intelligently adjusting the amplitude and phase shift of the incident signals. Furthermore, the authors of  \cite{9909807} studied RIS-aided multi-user multi-input single-output (MU-MISO) communications and target detection. To achieve wide-coverage and ultra-reliable communication performance in an ISAC scenario with incomplete self-interference cancellation, the authors of \cite{10364735} optimized the achievable sum-rate of multi-user communications. They also imposed a signal-to-interference-plus-noise ratio (SINR) constraint for target detection and a Cramer-Rao bound (CRB) constraint for parameter estimation, ensuring superior sensing performance for both tasks. Similarly,  RIS is deployed to mitigate self-interference and co-channel interference in non-orthogonal multiple access (NOMA) ISAC networks \cite{10423585}. The SINR of communication users and sensing targets were then optimized through the joint design of base station (BS) transmit beamforming, RIS phase shifts, and power allocation factors within each dynamic user cluster. To improve the performance of ISAC systems in imperfect channel state information (CSI) scenarios,  an RIS-assisted MIMO ISAC framework was proposed in \cite{10197455}, which deployed a distributed RIS consisting of a large-scale reflection sub-surface and two small-scale sensing sub-surfaces to achieve simultaneous communication and position sensing for blind spot users. \par
In the aforementioned RIS-assisted ISAC systems, the incident signal is only reflected by RIS, i.e., single-functional RIS (SF-RIS). Consequently, communication users, sensing targets, and the BS must be deployed on the same side of the RIS, limiting signal enhancement coverage to a half-space \cite{10411853}. To address this, the dual-functional RIS (DF-RIS) with reflection and refraction functions has been proposed, such as simultaneously transmitting and reflecting RIS (STAR-RIS) and  intelligent omni surface (IOS), which is deployed in ISAC systems to achieve full-space signal coverage \cite{10155669, 9965371,10464353}. A STAR-RIS-assisted ISAC system was first proposed in \cite{10050406}, which divides the entire space into a sensing space with a single target and a communication space with multiple users.  In addition, the authors of \cite{10178069} considered an ISAC system for single-target detection and multi-user communications, which is enabled by STAR-RIS with the energy splitting (ES) protocol. Although full-space signal enhancement can be achieved with the assistance of STAR-RIS, it may also pose serious security issues as all sensing targets within the signal coverage range may act as potential eavesdroppers \cite{10472878}. To prevent the eavesdropping while enhancing deployment flexibility,  \cite{10293862} considered a STAR-RIS-assisted ISAC system in the presence of an eavesdropper. The secrecy rate of the whole system is maximized by jointly optimizing the transmit beamforming and STAR-RIS coefficients. In addition, the echo from sensing targets needs to be reflected/refracted by STAR-RIS before reaching the BS in STAR-RIS-assisted ISAC systems. These multi-hop sensing links, namely BS-RIS-targets-RIS-BS links, experience significant path loss, which limits the sensing performance. To address this challenge, the authors in \cite{10464353} deployed a bi-static radar in an IOS-aided ISAC system to mitigate sensing path loss caused by multiple reflection/refraction hops, where receiving antennas are implemented in both the refraction and the reflection spaces to receive the echo signals. However, this design increases the hardware cost of the ISAC system due to the introduction of additional radar receiving equipment. 
 \begin{table*}[t] 
	\centering 
	\renewcommand\arraystretch{1.15}
	\caption{Comparison of this work with other representative works on RIS-aided ISAC systems}
	\scalebox{1.12}{
		\begin{tabular}{|c|c|c|c|c|c|c|c|c| } \hline 
			\multirow{3}{*}{\textbf{Ref.}}&\multicolumn{2}{c|}{\textbf{Scenarios}}&\multicolumn{2}{c|}{\textbf{Performance}}&\multicolumn{4}{c|}{\textbf{Supported Function}}  \\ \cline{2-9}
			&\makecell{Multiple\\users}&\makecell{Multiple\\ targets} &\makecell{Full-space\\ enhancement}&\makecell{Double-fading\\ mitigation}&Reflection&Refraction&Amplification&Sensing \\ \hline
			\cite{10478980}                        
			&&&\checkmark&\checkmark&\checkmark&\checkmark&& \checkmark             \\ \hline
			\cite{10274660, 10422722}      
			&\checkmark&&&&\checkmark&&&            \\ \hline
			\cite{10319318,10054402,9909807,10364735}
			&\checkmark&&&\checkmark&\checkmark&&\checkmark&              \\ \hline
			\cite{10423585,10411853}
			&\checkmark&\checkmark&&&\checkmark&&&              \\ \hline
			\cite{10155669}
			&\checkmark&\checkmark&\checkmark&&\checkmark&\checkmark&&\checkmark              \\ \hline
			\cite{10464353}
			&\checkmark&\checkmark&\checkmark&&\checkmark&\checkmark&&              \\ \hline
			\cite{10050406}
			&\checkmark&&\checkmark&&\checkmark&\checkmark&& \checkmark             \\ \hline
			\cite{10178069,10472878,10293862}
			&\checkmark&&\checkmark&&\checkmark&\checkmark&&             \\ \hline
			\cite{zhang2023joint}
			&\checkmark&&\checkmark&\checkmark&\checkmark&\checkmark&\checkmark&             \\ \hline
			\textbf{This work}
			&\checkmark&\checkmark&\checkmark&\checkmark&\checkmark&\checkmark&\checkmark&\checkmark             \\ \hline
	\end{tabular}}
	\label{T1}
\end{table*}
\subsection{Motivations and Challenges}
To enhance spectral efficiency and reduce hardware costs, it is highly meaningful to design a software and hardware resource-sharing system that enables ISAC for intelligent and diverse scenarios. Moreover, by providing a virtual LoS link, the introduction of RIS can effectively improve the sensing and communication performance of the ISAC system. However, the above works mainly focus on SF- and DF-RIS-assisted ISAC systems, which are limited in enhancing sensing and communication performances owing to the following reasons.
         \begin{itemize}
        	\item \textit{Firstly}, in conventional passive RIS- and active RIS-assisted ISAC systems, the enhanced coverage of signal transmission is limited to a half-space. To extend the enhanced coverage to a full-space, the coordinated operation of multiple RISs is required, which poses new challenges to the deployment and configuration of RISs.
        	\item \textit{Secondly}, although DF-RIS can achieve signal enhancement within a full-space coverage for ISAC systems, the double-fading effect introduced will cause attenuation of ISAC signals, especially for sensing signals passed through four-hop links.
        	\item \textit{Thirdly}, in ISAC systems, hardware costs will increase as the deployment of additional radars is necessary to receive the echo from sensing targets. 
          \end{itemize}\par
To overcome the above-mentioned issues faced by existing SF- and DF-RIS-assisted ISAC systems, we propose a multi-functional RIS (MF-RIS) architecture with simultaneous reflection, refraction, amplification, and self-sensing functions. These functions have been studied on the practical implementation, which provides a solid foundation to support the realization of the MF-RIS. In Table \ref{T1}, we compare existing studies with this work in terms of research scenarios, performance advantages, and RIS function supported.  By allowing reflection/refraction elements to flexibly switch between reflection/refraction modes, the MF-RIS achieves signal enhancement within a full-space coverage for communications and sensing, which also provides more DoFs for signal manipulation. In addition, the echos from the sensing target is received and processed by sensing elements of the MF-RIS, which effectively reduces the number of hops for sensing links, and the  double-fading effect can be compensated by the amplification function of the MF-RIS.
\subsection{Contributions }
While the architecture of MF-RIS has been proposed to support multiple functions such as signal reflection, refraction, amplification, and energy harvesting, existing studies have not integrated the sensing function into MF-RIS \cite{10198355, 10225701, 10021586, 10214569}. To verify the feasibility of the proposed MF-RIS with self-sensing function in ISAC systems, we  investigate an echo SINR maximization problem in an MF-RIS-enabled ISAC network, where a multi-user and multi-target scenario is considered. The main contributions of this paper are summarized as follows: 
    \begin{itemize}
      	\item We propose an MF-RIS enabled ISAC system to achieve full-space enhanced coverage for communication and sensing signals.  To endow MF-RIS with the self-sensing function, low-cost sensing elements are deployed on MF-RIS to replace traditional radars to receive the echo from targets and extract angle information of targets.  We consider a more practical and general scenario where multiple users and multiple targets are distributed throughout the entire space, which is divided by MF-RIS into two half-spaces, namely the refraction space and the reflection space.
     	\item We formulate an optimization problem to maximize the SINR of echo from multiple targets by jointly optimizing the transmit beamforming, receiving vectors of sensing elements, and reflection/refraction element coefficients. To solve this non-convex problem, we decouple the original problem into three sub-problems, i.e., sensing filter design, transmit beamforming optimization, and MF-RIS coefficients design. Then, an alternating optimization (AO) algorithm is proposed to iteratively solve them.
	    \item The superiority of the MF-RIS in ISAC systems and the convergence of the proposed algorithm are demonstrated by extensive simulation. The numerical results show that:  1) The proposed MF-RIS with the ES protocol exhibits the best performance compared with the  mode switching (MS) and time switching (TS) protocols; 2) The proposed MF-RIS effectively improves sensing performance of ISAC systems while excellent communication performance is retained, compared to active RIS, passive RIS, and STAR-RIS-assisted ISAC systems.
      \end{itemize}\par
\begin{table}[t] 
		\renewcommand\arraystretch{1.2}
	\centering 
	\caption{List of Notations}
		\scalebox{1}{
		\begin{tabular}{|c|c|} \hline 
			\bf{Notations} & \bf{Definition} \\ \hline		
			$N$ & Number of transmit antennas \\ \hline  
			$M$ & Number of reflection/refraction elements \\ \hline 
			$M_s$ & Number of sensing elements \\ \hline 
			$\mathcal{D}$& Set of spaces \\ \hline
			$\mathcal{K}_d$& Set of users in space $d$\\ \hline
			$\mathcal{J}_d$ &Set of targets in space $d$ \\ \hline
			$K_d$ & Number of users in space $d$\\ \hline 
			$J_d$ & Number of targets in space $d$\\ \hline 
			$L$ & Number of time indexes\\ \hline
			$\kappa$ & Rician factor\\ \hline
			$\lambda$ & Carrier wavelength \\\hline
			$\zeta$ & Antenna or element separation \\ \hline
			$\mathbf{H}$ & Channel between the BS and the MF-RIS \\ \hline 
			$\mathbf{g}_{k}$ &Channel between the MF-RIS and the $k$-th user \\ \hline 
			$\mathbf{G}_d$ & Multi-target response matrix in space $d$  \\ \hline 
			$\sigma_k^2$ & Noise power for the $k$-th user\\ \hline 
			$\sigma_r^2$ & Noise power for reflection/refraction elements \\ \hline 
			$\sigma_s^2$ & Noise power for sensing elements \\ \hline 
			$\gamma_k$ & Receiving SINR of the $k$-th user \\ \hline
			$\gamma_d$ & Echo SINR received by the MF-RIS in space $d$   \\ \hline
			$\gamma_{\rm{th}}$ &QoS requirement of each user \\ \hline
			$P^{\rm{max}}_{\rm{total}}$& Total power budget of the system \\ \hline
			$P^{\rm{max}}_{\rm{b}}$ &Maximum transmit power of the BS\\ \hline
			$P^{\rm{max}}_{\rm{r}}$& Power budget of the MF-RIS \\ \hline
			$\mathcal{R}_{\rm{MF}}$ & Feasible set of the MF-RIS coefficients \\ \hline
	\end{tabular}}
	\label{T2}
\end{table}
The rest of this paper is organized as follows. Section \ref{S2} presents the proposed MF-RIS-enabled ISAC model and the formulated problem. The echo maximization problem is solved in Section \ref{S3}. Numerical results are revealed in Section \ref{S4}, followed by conclusions in Section \ref{S5}.
\section{System Model and Problem Formulation}\label{S2}
\begin{figure*}[t]
	\centering{}\includegraphics[width=7.5in]{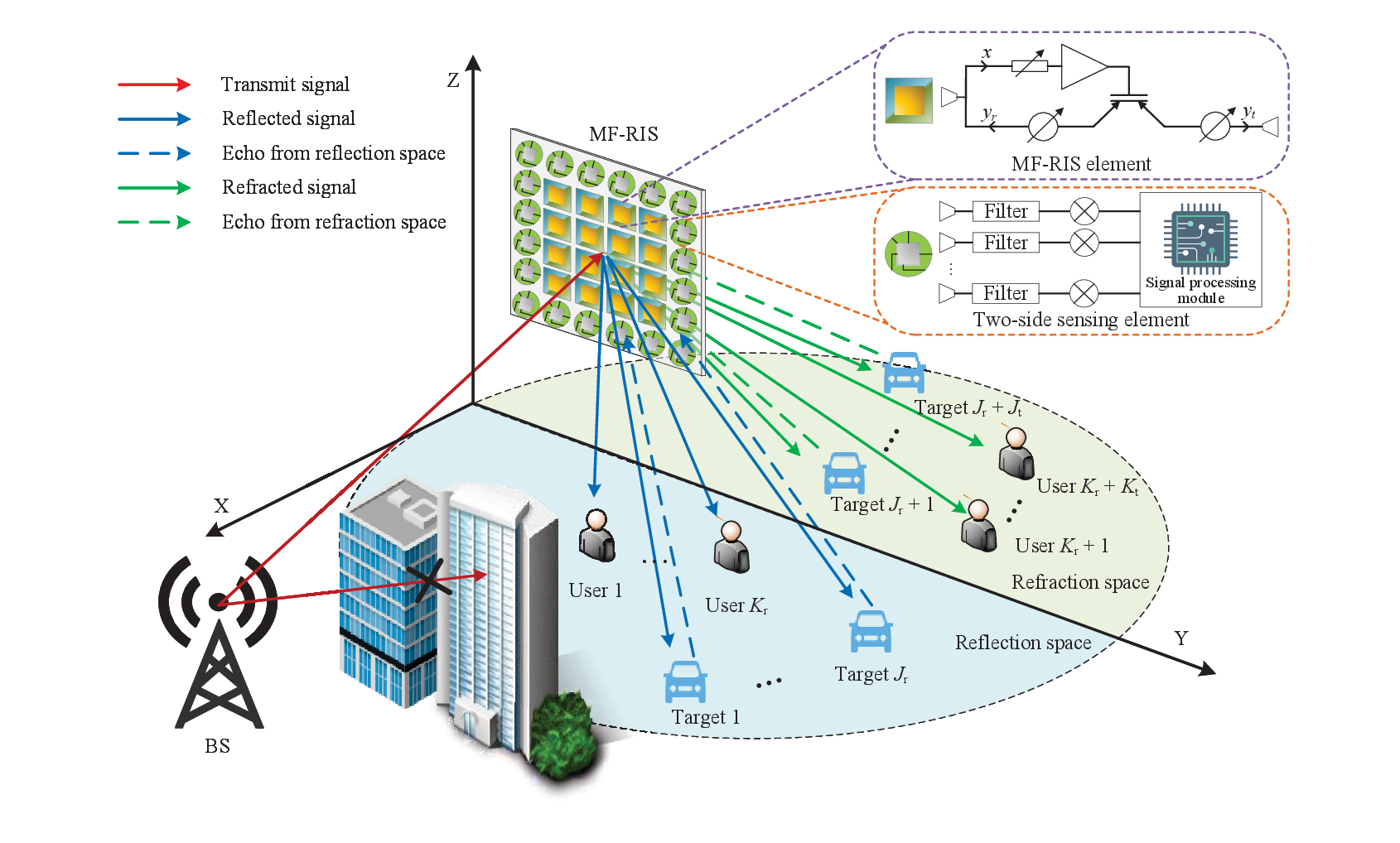}
	\caption{Illustration of the proposed MF-RIS-enabled multi-user and multi-target ISAC system.}
	\label{system_model}
\end{figure*}
As illustrated in Fig. \ref{system_model}, we consider an MF-RIS enabled  ISAC network, where an $N$-antenna BS is deployed to provide downlink communication service for $K$ single-antenna users. The MF-RIS consists of $M$ reflection/refraction elements and $M_s$ two-side sensing elements, which are used to reflect/refract ISAC signals and support the sensing function for detecting $J$ targets.\footnote{The sensing elements are a dedicated low-cost sensor, which have lower hardware cost than the antennas at the BS \cite{10443321}. } The MF-RIS divides the ISAC signal coverage space into two half-spaces, namely the reflection space $r$, and the refraction space $t$,  thus the space set can be indexed by $\mathcal{D}=\left\{r, t\right\}$. There are $K_r$ single-antenna communication users and $J_r$ sensed targets in the reflection space $r$, and $K_t$ single-antenna communication users and $J_t$ sensed targets in the refraction space $t$, which are denoted by the sets $\mathcal{K}_r=\left \{1,...,K_r\right\}$, $\mathcal{J}_r=\left\{1,...,J_r\right\}$, $\mathcal{K}_t=\left \{K_r+1,...,K_r+K_t\right\}$, and $\mathcal{J}_t=\left\{J_r+1,...J_r+J_t\right\}$, satisfying $K_r+K_t=K$ and $J_r+J_t=J$, respectively. To simplify the notations, user and target sets in half-space $d$ are denoted by $\mathcal{K}_d$ and $\mathcal{J}_d$, respectively. Note that when $d=r$, $k\in\mathcal{K}_r$, $j\in\mathcal{J}_r$; when $d=t$, $k\in\mathcal{K }_t$ and $j\in\mathcal{J}_t$.\par
In the proposed MF-RIS-enabled ISAC system, we assume that the LoS links between the BS and users/targets are occluded. Furthermore, we consider a quasi-static channel model with the coherent time block of length $L$, during which the communication channels and sensing target parameters remain approximately constant \cite{10050406,9531484,9652071}.\footnote{The time-varying channel model is also commonly used in ISAC systems to detect the velocity of moving targets by analyzing the Doppler shift \cite{10478980}. However, this topic is beyond the scope of this paper, and we will further utilize a time-varying channel model to address mobility-related challenges in the future.} At the time index $ l\in\{1,2,...,L\}$, the BS transmits communication data streams $\mathbf{c}(l)$ to $K$ users simultaneously, and the data stream carrying the $k$-th user information is denoted by $c_k(l)$. To ensure the sensing performance of transmit signals, additional sensing streams $\mathbf{s}(l)$ are introduced as dedicated sensing signals. Then, multi-target sensing and multi-user communication in full-space signal enhanced converge can be achieved with the aid of the MF-RIS. Therefore, the transmit signal of the BS can be expressed as
\begin{align}
	\mathbf{x}(l)=\mathbf{W}_c\mathbf{c}(l)+\mathbf{F}_s\mathbf{s}(l)=\sum_{d\in\mathcal{D}}\sum_{k\in\mathcal{K}_d}\mathbf{w}_kc_k(l)+\sum_{n\in\mathcal{N}}\mathbf{f}_ns_n(l),
\end{align}
where $\mathcal{N}=\left\{1,2,...,N\right\}$ is the set of BS antennas. $\mathbf{W}_c=\left[\mathbf{w}_1,...,\mathbf{w}_{K_r},\mathbf{w}_{K_r+1},...,\mathbf{w}_{K_r+K_t}\right]\in\mathbb{C}^{N\times K}$ represents the beamforming matrix used to deliver data streams  $\mathbf{c}(l)=\left[c_1(l),c_2(l),...,c_K(l)\right]^T\in\mathbb{C}^{K\times1}$ to $K$ communication users. $\mathbf{s}(l)=\left[s_1(l),s_2(l),...,s_N(l)\right]^T$ is a dedicated sensing signal which includes $N$ independent radar waveforms, and $\mathbf{F}_s=\left[\mathbf{f}_1,\mathbf{f}_2,...,\mathbf{f}_n\right]$ is its associated beamformer. Furthermore, the communication signal is modeled as an independent Gaussian random signal with zero mean and unit power, while the dedicated sensing signal is generated by pseudo-random coding \cite{10050406}. We can then assume that communication symbols for different users are uncorrelated, individual radar waveforms are uncorrelated with each other and communication symbols are uncorrelated with radar waveforms, i.e., $\mathbb{E}\left[\mathbf{c}(t )\mathbf{c}(l)^H\right]=\mathbf{I}_K$ , $\mathbb{E}\left[\mathbf{s}(l)\mathbf{s}(l)^H \right]=\mathbf{I}_N$, and $\mathbb{E}\left[\mathbf{c}(l)\mathbf{s}(l)^H\right]=\mathbf{0}_{ K\times M}$.
\subsection{MF-RIS Model}\label{2A}
In a three-dimensional (3D) Cartesian coordinate system, we consider that the proposed MF-RIS is a uniform planar array (UPA) parallel to the $Y$-$Z$ plane, which consists of  $M=M_y\times M_z$ reflection/refraction elements. In addition,  the MF-RIS is equipped with $M_v$ vertical sensing elements and $M_h$ horizontal sensing elements that satisfy $M_v+M_h=M_s$, which are used to estimate the elevation and azimuth angles from the echo of sensing targets. Define $x_m$ as the signal received by the $m$-th reflection/reflection element, then the signal after being reflected and refracted by the $m$-th reflection/reflection elements can be expressed as $y_m^r=\sqrt{\beta_m^r}e^{\theta_m^r}$ and $y_m^t=\sqrt{\beta_m^t}e^{\theta_m^t}$, where $\theta_m^r,\theta_m^t$ and $\beta_m^r,\beta_m^t$ represent reflection and refraction phase shifts, as well as reflection and refraction amplitude coefficients of the $m$-th element, respectively.  The reflection and refraction coefficients of MF-RIS elements can be expressed as  
\begin{align}
	\mathbf{\Theta}_r=\mathrm{diag}\left( \bm{\theta}_r\right),
	\mathbf{\Theta}_t=\mathrm{diag}\left( \bm{\theta}_t\right),
\end{align}
where $\bm{\theta}_r = [\sqrt{\beta_1^r}e^{j\theta_1^r},\sqrt{\beta_2^r}e^{j\theta_2^r},...,\sqrt{\beta_M^r}e^{j\theta_M^r}]^H\in\mathbb{C}^{M\times 1}$ and $\bm{\theta}_t = [\sqrt{\beta_1^t}e^{j\theta_1^t},\sqrt{\beta_2^t}e^{j\theta_2^t},...,\sqrt{\beta_M^t}e^{j\theta_M^t}]^H\in\mathbb{C}^{M\times 1}$ are reflection and reflection beamforming vectors, respectively, which satisfy $\theta_m^r, \theta_m^t\in\left[ 0,2\pi\right) $. Note that the amplifier cannot consume more energy than the maximum available energy that the MF-RIS can provide, i.e. , $\beta_m^t, \beta_m^r \in \left[ 0, \beta_{\max}\right] $ and $\beta_m^t+\beta_m^r\le\beta_{\max}$.
\subsection{Communication Model}
To characterize the maximum performance potential of the MF-RIS enabled ISAC system, we assume that perfect CSI  has been obtained for all channels using the channel estimation method similar to \cite{9864300}. In the proposed system model, both LoS and non-LoS (NLoS) links exist as the BS and the MF-RIS are generally set at relatively high altitudes. Therefore, we adopt Rician fading to model all channels, which are assumed to be uncorrelated as \cite{10254508,zhang2023joint}. Thus, the channel matrix $\mathbf{H}\in\mathbb{C}^{M\times N}$ between the BS and the MF-RIS can be expressed as 
\begin{align}
	\mathbf{H}=\sqrt{L_{br}}\left(\sqrt{\frac{\kappa}{\kappa+1}}\mathbf{H}^{\mathrm{LoS}}+\sqrt{\frac{1}{\kappa+1}}\mathbf{H}^{\mathrm{NLoS}}\right), \label{H}
\end{align}%
where $\kappa$ is the Rician factor and $L_{br}$ is the distance-dependent path loss, which can be further defined as $L_{br}=h_0d^{-\alpha_0}_{br}$.  The parameters $h_0$, $d_{br}$ and $\alpha_0$ represent the path loss at a reference distance of 1 meter (m), the link distance from the BS to the MF-RIS, and the corresponding path loss  exponent, respectively. The matrix $\mathbf{H}^{\mathrm{NLoS}}$ represents the NLoS component, which obeys the Rayleigh fading model with  independent and identically distributed (i.i.d.) elements. Define $\mathbf{u}(\widetilde{\theta},\widetilde{N})$ as a one-dimensional steering vector function, which is defined as follows \cite{9652071}:
\begin{align}
	\mathbf{u}(\widetilde{\theta},\widetilde{N})=\left[1,e^{-j2\pi\widetilde{\theta}},...,e^{-j2\pi(\widetilde{N}-1)\widetilde{\theta}}\right]^T,
\end{align}
where $\widetilde{\theta}$ represents the phase difference between adjacent antennas/elements, and $\widetilde{N}$ represents the number of antennas/elements. Therefore, the LoS component matrix $\mathbf {H}^{\mathrm{LoS}}$ can be expressed as
\begin{align}
	\mathbf{H}^{\mathrm{LoS}}=&\mathbf{u}\left(\frac{\zeta\sin{\varphi_{br}}}{\lambda},M_z\right)
	\otimes\mathbf{u}\left(\frac{\zeta\cos{\varphi_{br}}\cos{\phi_{br}}}{\lambda},M_y\right)\nonumber \\
	&\otimes\mathbf{u}^T\left(\frac{\zeta\cos{\varphi_{br}}\cos{\phi_{br}}}{\lambda},N\right),
\end{align}
where $\zeta$ and $\lambda$  represent the antenna/element separation and the carrier wavelength, respectively, and the symbol $\otimes$ represents the Kronecker product. $\varphi_{br}$ and $\phi_{br}$ are used to describe the elevation angle and the azimuth angle of the MF-RIS relative to the BS, respectively. We denote the channel vector from the MF-RIS to the $k$-th user by $\mathbf{g}_{k}\in\mathbb{C}^{M\times 1}$ that is obtained through a generation mechanism similar to $\mathbf {H}$ in \eqref{H}, and is given by
\begin{align}
	\mathbf{g}_{k}=\sqrt{h_0d^{-\alpha_k}_{rk}}\left(\sqrt{\frac{\kappa}{\kappa+1}}\mathbf{g}^{\mathrm{LoS}}_{k}+\sqrt{\frac{1}{\kappa+1}}\mathbf{g}^{\mathrm{NLoS}}_{k}\right).
\end{align}\par
Then, the received signal of the $k$-th user in space $d$ can be expressed as
\begin{align}
	y_{k}(l)=\mathbf{g}_{k}^H\mathbf{\Theta}_d\mathbf{H}\mathbf{x}(l)+\mathbf{g}_{k}^H\mathbf{\Theta}_d\mathbf{n}_r(l)+n_{k}(l),\forall k\in\mathcal{K}_d,
\end{align}
where $\mathbf{n}_r(l)\sim\mathcal{CN}(\mathbf{0},\sigma^2_r\mathbf{I}_M)$ represents the amplification noise generated by the MF-RIS with per-element noise power $\sigma^2_r$. $n_{k}(l)\sim\mathcal{CN}(0,\sigma^2_k)$ represents the additive Gaussian white noise (AWGN) received by the $k$-th user in space $d$ with power $\sigma^2_k$. For conventional passive RISs, $\mathbf{g}_k^H\mathbf{\Theta}_d\mathbf{n}_r$ is very small compared to $n_k$, and can be ignored \cite{10225701}. However, such a noise cannot be ignored since it is amplified by the MF-RIS. Then, the achievable communication rate of the $k$-th user in space $d$ can be expressed as
\begin{align}
	R_{k}=\log_2(1+\gamma_{k}),\forall k\in\mathcal{K}_d, \forall d\in\mathcal{D},
\end{align}
where $\gamma_{k}$  represents the SINR of the $k$-th user in space $d$. The expression of $\gamma_k$ is given by \eqref{SINR} at the top of next page. 
\begin{figure*}[t]
	\begin{equation}
		\gamma_{k}=\frac{\left|\mathbf{g}_{k}^H\mathbf{\Theta}_d\mathbf{H}\mathbf{w}_k\right|^2}{\sum_{i\in\mathcal{K},i\neq k}\left|\mathbf{g}_{k}^H\mathbf{\Theta}_d\mathbf{H}\mathbf{w}_i\right|^2+\sum_{n\in\mathcal{N}}\left|\mathbf{g}_{k}^H\mathbf{\Theta}_d\mathbf{H}\mathbf{f}_n\right|^2+\sigma^2_r\left\| \mathbf{g}_{k}^H\mathbf{\Theta}_d\right\| ^2+\sigma^2_{k}},\forall k\in\mathcal{K}_d, \forall d\in\mathcal{D}. \label{SINR}
	\end{equation}
	\hrulefill
\end{figure*}
\subsection{Sensing Model}
In the considered MF-RIS enabled ISAC system, $M_s$ sensing elements are deployed in the MF-RIS to receive the echos from $J$ sensing targets for estimating  channel parameters. Since the direct links between the BS and targets are obscured, we consider only the BS-RIS-target-RIS links in the process of sensing. Specifically, the BS transmits ISAC signals with dedicated sensing streams to the MF-RIS, which are then amplified and transmitted to reflection/refraction space by the MF-RIS reflection/refraction elements. Additionally, the MF-RIS sensing elements are employed to receive the echo from targets. Note that there is no need to consider the echo interference from the opposite half-space due to the use of two-side sensors \cite{10050406,singh2023multi}. \par
In the $l$-th time index, the echo signal received by the MF-RIS sensing elements in space $d$ can be modeled as 
\begin{align}
	\mathbf{y}_d(l)=\sum_{j\in\mathcal{J}_d}\alpha_{j}\mathbf{a}(\varphi_{j},\phi_{j})\mathbf{b}^T(\varphi_{j},\phi_{j})\mathbf{\Theta}_d\mathbf{H}\mathbf{x}(l)+\mathbf{n}_{d}(l),  \label{y_dt}
\end{align}
where $\alpha_j$ is a complex coefficient and represents the round-trip path loss of the echo signal from the $j$-th target, which is related to radar cross-section (RCS). $\mathbf{n}_d\sim\mathcal{CN}(\mathbf{0},\sigma_s^2\mathbf{I}_{M_s})$ represents the noise received by the sensing elements in space $d$ with per-element power $\sigma_s^2$. The parameters $\varphi_{j}$ and $\phi_{j}$ are the elevation and azimuth angle of the $j$-th target with respect to the MF-RIS, respectively. $\mathbf{b}(\varphi_{j},\phi_{j})$ and $\mathbf{a}(\varphi_{j},\phi_{j})$ represent the steering vectors from the MF-RIS to the target and the target to the MF-RIS, respectively, which are given by 
\begin{align}
	&\mathbf{b}(\varphi_{j},\phi_{j})=\mathbf{u}\left(\frac{\zeta\sin{\varphi_{j}}}{\lambda},M_z\right)
	\otimes\mathbf{u}\left(\frac{\zeta\cos{\varphi_{j}}\cos{\phi_{j}}}{\lambda},M_y\right),\\
	&\mathbf{a}(\varphi_{j},\phi_{j})=\left[\mathbf{u}\left(\frac{\zeta\sin{\varphi_{j}}}{\lambda},M_v\right); \mathbf{u}\left(\frac{\zeta\cos{\varphi_{j}}\cos{\phi_{j}}}{\lambda},M_h\right)\right].
\end{align}
In a coherent time block of length $L$, according to \eqref{y_dt}, the received echo signal from the target at the sensing elements can be expressed as
\begin{align}
	\mathbf{Y}_d=\mathbf{A}\mathbf{\alpha}\mathbf{B}^H\mathbf{\Theta}_d\mathbf{H}\mathbf{X}+\mathbf{N}_s,\label{Y_d}
\end{align}
where 
\begin{subequations}
	\begin{align}
		&\mathbf{X} = \left[ \mathbf{x}(1),\mathbf{x}(2), ... ,\mathbf{x}(L)\right], \\
		&\mathbf{Y}_d = \left[ \mathbf{y}_d(1),\mathbf{y}_d(2),...,\mathbf{y}_d(L)\right], \\
		&\mathbf{A}  = \left[ \mathbf{a}(\varphi_{1},\phi_{1}), \mathbf{a}(\varphi_{2},\phi_{2}) ,..., \mathbf{a}(\varphi_{J_d},\phi_{J_d})\right], \\
		&\mathbf{\alpha}=\mathrm{diag} \left(\alpha_{1},\alpha_{2},...,\alpha_{J_d}\right),\\
		&\mathbf{B} = \left[ \mathbf{b}(\varphi_{1},\phi_{1}),\mathbf{b}(\varphi_{2},\phi_{2}),...,\mathbf{b}(\varphi_{J_d},\phi_{J_d})\right], \\
		&\mathbf{N}_s = \left[ \mathbf{n}_s(1),\mathbf{n}_s(2),...,\mathbf{n}_s(L)\right]. 
	\end{align}
\end{subequations}\par
In the initial phase for target detection, there is no prior knowledge of the target. For this case, the MF-RIS aims to estimate the complete multi-target response matrix $\mathbf{G}_d=\mathbf{A }\mathbf{\alpha}\mathbf{B}^H$ based on the received signal $\mathbf{Y}_d$ in \eqref{Y_d}, or equivalently estimate $\mathbf{\vec{g}}_d$ based on $\mathbf{\vec{y}}_d$, which is given by
\begin{align}
	\mathbf{\vec{y}}_d=\mathrm{vce}(\mathbf{G}_d\mathbf{\Theta}_d\mathbf{H}\mathbf{X}+\mathbf{N}_s)=(\mathbf{X}^H\otimes\mathbf{I}_{M_s})\mathbf{\vec{g}}_d+\mathbf{\tilde{n}}_d\label{y_d},
\end{align}
where $\mathbf{\vec{y}}_d=\mathrm{vec}(\mathbf{Y}_d)$, $\mathbf{\vec{g}}_d=\mathrm{vec}(\mathbf{G}_d\mathbf{\Theta}_d\mathbf{H})$, and $\mathbf{\vec{n}}_d=\mathrm{vec}(\mathbf{N}_d)$.  It can be observed from \eqref{y_d} that the received signal vector $\mathbf{\vec{y}}_d$ is a circularly symmetric complex Gaussian (CSCG) random vector with a mean value of  $(\mathbf{X}^H\otimes\mathbf{I}_ {M_s})\mathbf{\tilde{g}}_d$ and a variance of $\sigma^2_s\mathbf{I}_{M_s}$, i.e.,  $\mathbf{\vec{y}}_d\sim\mathcal {CN}((\mathbf{X}^H\otimes\mathbf{I}_{M_s})\mathbf{\tilde{g}}_d,\sigma^2_s\mathbf{I}_{M_s})$. According to the complex linear model in \eqref{y_d}, the MF-RIS needs to estimate $M_sN$ complex parameters in $\mathbf{G}_d$ or $\mathbf{\tilde{g}}_d$. The CRB matrix for estimating $\mathbf{G}_d$ has been determined in \cite{10217169}.\par
Based on the estimated value of $\mathbf{G}_d$, some valuable sensing information can be extracted, including the target’s location via time delay analysis,  angle via  direct-of-arrival (DoA) estimation algorithm, as well as shape, material and size via RCS analysis. For example, MUSIC (Multiple Signal Classification) is a classical high-resolution DoA estimation algorithm, which is particularly effective in multi-target scenarios \cite{9724202}. The accuracy of these estimation algorithms and parameter analysis heavily relies on the sensing SINR, as the probability of target detection is typically a monotonically increasing function of the sensing SINR \cite{10464353,zhang2023joint,10319318}. A higher sensing SINR reduces the impact of noise and interference, meaning that the radar can more easily distinguish and extract target information from echo signal, thereby enhancing detection accuracy. Therefore, the echo SINR from sensed target is an essential and widely used performance metric to evaluate the sensing performance of the MF-RIS. \par
\begin{figure*}[t]
	\begin{equation}
		\gamma_{d}=\frac{\sum_{k\in\mathcal{K}}\left| \mathbf{m}_d^H\mathbf{G}_d\mathbf{\Theta}_d\mathbf{H}\mathbf{w}_k\right|^2+ \sum_{n\in\mathcal{N}}\left|\mathbf{m}_{d}^H\mathbf{G}_{d}\mathbf{\Theta}_d\mathbf{H}\mathbf{f}_n\right|^2}{\sum_{k\in\mathcal{K}}\left|\mathbf{m}_{d}^H\mathbf{H}_d\mathbf{w}_k\right|^2+\sum_{n\in\mathcal{N}}\left|\mathbf{m}_{d}^H\mathbf{H}_d\mathbf{f}_n\right|^2+\sigma^2_r\left\| \mathbf{m}_{d}^H\mathbf{G}_d\mathbf{\Theta}_d\right\| ^2+\sigma^2_s\left\| \mathbf{m}_{d}^H\mathbf{I}_{M_s}\right\| ^2},  \forall d\in\mathcal{D}. \label{SINR_d}
	\end{equation}
	\hrulefill
\end{figure*}
We denote the channel from the BS to the sensing elements in space $d$ by $\mathbf{H}_{d}$. Then, the SINR of echo received by the sensing elements in space $d$ is given by \eqref{SINR_d} at the top of next page.
\subsection{Problem Formulation}
In the considered MF-RIS enabled ISAC system, it is essential to ensure that sensing elements can effectively detect and extract the echo signal while meeting the quality of service (QoS) requirements of all communication users. The sensing performance is determined by the SINR of the echo signal from the target received by the sensing elements. Therefore, our goal is to maximize the SINR in both sub-space $t$ and sub-space $r$, i.e., $\max \gamma_r +\max \gamma_t$, by jointly optimizing the transmit beamforming $\left\{ \mathbf{w}_k ,\mathbf{f}_n\right\} $,  reflection/refraction element coefficients $\left\{\mathbf{\Theta}_r, \mathbf{\Theta}_t\right\}$, and MF-RIS sensing filter $\left\{\mathbf{m}_r, \mathbf{m}_t\right\}$. Since the MF-RIS employs two-sided sensing elements to divide the entire space into two uncorrelated sub-spaces,  the sum of maximizing the sensing SINR can be equivalently replaced by maximizing the sum of  the sensing SINR, i.e., $\max \gamma_r +\max \gamma_t= \max\sum_{d\in\{r,t\}}\gamma_d$. Thus, the optimization problem is formulated as
\begin{subequations}
	\begin{align}
		\underset{\mathbf{w}_k, \mathbf{f}_n,\mathbf{\Theta}_d, \mathbf{m}_{d}}{\max} \quad   
		&\sum_{d\in \mathcal{D}}\gamma_d \label{P0a}\\
		\mathrm{s.t.}\quad \quad
		&\sum_{k\in\mathcal{K}}\left\| \mathbf{w}_k\right\| ^2+\sum_{n\in\mathcal{N}} \left\| \mathbf{f}_n\right\| ^2  \le P^{\max}_{\rm{b}}\label{P0b},\\
		&R_{k} \ge R_{\rm{th}}, \forall k\in\mathcal{K}_d,\forall d\in\mathcal{D},\label{P0c}\\
		&\beta_m^d,\theta_m^d \in \mathcal{R}_{\rm{MF}},\forall m, \forall d,\label{P0d}\\
		&\sum_{d\in\mathcal{D}}\sum_{k\in\mathcal{K}}\left\| \mathbf{\Theta}_d\mathbf{H}\mathbf{w}_k\right\| ^2+\sum_{d\in\mathcal{D}}\sum_{n\in\mathcal{N}}\left\| \mathbf{\Theta}_d\mathbf{H}\mathbf{f}_n\right\| ^2\nonumber\\
		&+\sigma_r^2\sum_{d\in\mathcal{D}}\left\|\mathbf{\Theta}_d \right\| ^2 \le P^{\max}_{\rm{r}},\label{P0e}\\
		&\left\|\mathbf{m}_t \right\|^2 = \left\|\mathbf{m}_r \right\|^2=P_{\rm{s}} ,\label{P0f}
	\end{align} \label{P0}%
\end{subequations}%
where constraint \eqref{P0b} indicates that the maximum transmission power is limited by the transmit power budget $P^{\max}_{\rm{b}}$, and constraint \eqref{P0c} ensures that the achievable communication rate of the $k$-th user is above the QoS requirement $R_{\rm{th}}$.  The feasible set $\mathcal{R}_{\rm{MF}}$ in constraint \eqref{P0d} is given by $\mathcal{R}_{\rm{MF}}= \{ \beta_m^d, \theta_m^d | , \beta_m^d \in \left[ 0, \beta_{\max}\right], \sum_{d\in\mathcal{D}}\beta_m^d \le \beta_{\max}, \theta_m^d\in\left[0,2\pi \right),\forall m,d \} $.
\eqref{P0e} is the MF-RIS amplification power constraint, where $P^{\max}_{\rm{r}}$ denotes the power budget of the MF-RIS. $\mathbf{m}_t$ and $\mathbf{m}_r$ represent the filter when receiving the echo from the sub-space $t$ and the sub-space $t$, respectively, and constraint \eqref{P0f} means that the sensing elements receive the echo signal at a constant power $P_{\rm{s}}$. \par
Problem \eqref{P0} is difficult to solve directly due to the following challenges: 1) The objective function \eqref{P0a} and constraint \eqref{P0c} are non-convex, each of which involves complex logarithmic or fractional terms; 2) Optimization variables are highly coupled in the objective function \eqref{P0a}, constraints \eqref{P0c} and \eqref{P0e}.

\section{Problem Solution}\label{S3}
Due to the non-convex objective function \eqref{P0a} and constraints \eqref{P0b}, \eqref{P0c}, and \eqref{P0e},  we can not directly solve the original problem \eqref {P0}. To effectively solve the SINR maximization problem, we decouple it into three sub-problems, i.e., the sensing filter design sub-problem, the transmit optimization sub-problem, and the MF-RIS coefficient design sub-problem. Then, we propose an AO algorithm to iteratively solve them \cite{9681707}. Specifically, we obtain a closed-form expression for the optimal sensing filter using the eigenvalue decomposition method. For the transmit beamforming optimization and the MF-RIS coefficient design, the fractional programming method is used to reformulate these sub-problems, and then successive convex approximation (SCA) and second-order cone programming (SOCP) methods are adopted to obtain their local optimal solutions. 
\subsection{Sensing Filter Design}
To start with, we focus on optimizing receiving vectors $\{\mathbf{m}_{t},\mathbf{m}_{r}\}$ of sensing elements, given the transmit beamforming $\left\{ \mathbf{w}_k, \mathbf{f}_n \right\}$ and the MF-RIS reflection/refraction coefficients $\left\{\mathbf{\Theta}_r, \mathbf{\Theta}_t\right\}$. The sensing filter design problem is formulated as
\begin{subequations}
	\begin{align}
		\underset{\mathbf{m}_{t},\mathbf{m}_{r}}{\max} \quad   
		&\sum_{d\in \mathcal{D}}\gamma_d  \label{P1a}\\
		\mathrm{s.t.}\quad
		&\left\|\mathbf{m}_t \right\|^2 = \left\|\mathbf{m}_r \right\|^2=P_{\rm{s}}.\label{P1b}
	\end{align}\label{P1}%
\end{subequations}
Problem \eqref{P1} is non-convex, as the objective function \eqref{P1a} contains fractional terms involving variables $\mathbf{m}_r$ and $\mathbf{m}_t$. 
\begin{figure*}[t]
	\begin{equation}
		\gamma_{d}=\frac{\mathbf{m}_d^H \mathbf{G}_d \mathbf{\Theta}_d \mathbf{H}\left(\sum_{k\in\mathcal{K}}\mathbf{w}_k\mathbf{w}_k^H+\sum_{n\in\mathcal{N}}\mathbf{f}_n\mathbf{f}_n^H \right)\mathbf{H}^H\mathbf{\Theta}_d^H\mathbf{G}_d^H\mathbf{m}_d }{\mathbf{m}_d^H\left( \mathbf{H}_d\left(\sum_{k\in\mathcal{K}}\mathbf{w}_k\mathbf{w}_k^H+\sum_{n\in\mathcal{N}}\mathbf{f}_n\mathbf{f}_n^H \right)\mathbf{H}_d^H+\sigma_r^2\mathbf{G}_d\mathbf{\Theta}_d\mathbf{\Theta}_d^H\mathbf{G}^H+\sigma_s^2\mathbf{I}_{M_s}\right)\mathbf{m}_d }, \forall d\in\mathcal{D}. \label{S_d}
	\end{equation}
	\hrulefill
\end{figure*}%
Based on the assumption that the communication symbols of each user and the radar symbols of each antenna are uncorrelated, the SINR of the target echo received by the MF-RIS sensing elements in $d$ space is equivalent to \eqref{S_d} at the top of this page. To facilitate the solution of problem \eqref{P1}, we define 
\begin{align}
	&\tilde{\mathbf{H}}_d=\mathbf{G}_d \mathbf{\Theta}_d \mathbf{H}\left(\sum_{k\in\mathcal{K}}\mathbf{w}_k\mathbf{w}_k^H+\sum_{n\in\mathcal{N}}\mathbf{f}_n\mathbf{f}_n^H \right)\mathbf{H}^H\mathbf{\Theta}_d^H\mathbf{G}_d^H,\\
	&\hat{\mathbf{H}}_d=\mathbf{H}_d\left(\sum_{k\in\mathcal{K}}\mathbf{w}_k\mathbf{w}_k^H+\sum_{n\in\mathcal{N}}\mathbf{f}_n\mathbf{f}_n^H \right)\mathbf{H}_d^H+\bm{\sigma},
\end{align}
where $\bm{\sigma}=\sigma_r^2\mathbf{G}_d\mathbf{\Theta}_d\mathbf{\Theta}_d^H\mathbf{G}^H+\sigma_s^2\mathbf{I}_{M_s}$.
Then, we have 
\begin{align}
	\gamma_{d}=\frac{\mathbf{m}_d^H\tilde{\mathbf{H}}_d\mathbf{m}_d}{\mathbf{m}_d^H\hat{\mathbf{H}}_d\mathbf{m}_d}, \forall d\in\mathcal{D}.
\end{align}
Problem \eqref{P1} is equivalent to 
\begin{subequations}
	\begin{align}
		\underset{\mathbf{m}_{r},\mathbf{m}_{t}}{\max}  \quad   
		&\frac{\mathbf{m}_r^H\tilde{\mathbf{H}}_r\mathbf{m}_r}{\mathbf{m}_r^H\hat{\mathbf{H}}_r\mathbf{m}_r}+\frac{\mathbf{m}_t^H\tilde{\mathbf{H}}_t\mathbf{m}_t}{\mathbf{m}_t^H\hat{\mathbf{H}}_t\mathbf{m}_t}  \label{P1.1a}\\
		\mathrm{s.t.}\quad
		&\left\|\mathbf{m}_r \right\|^2=\left\|\mathbf{m}_t \right\|^2 =P_{\rm{s}} .\label{P1.1b}
	\end{align}\label{P1.1}%
\end{subequations}
Since variables $\mathbf{m}_r$ and $\mathbf{m}_t$ are not coupled in problem \eqref{P1.1}, the objective function \eqref{P1.1a} is equivalent to
\begin{align}
	&\underset{\mathbf{m}_{r}+\mathbf{m}_{t}}{\max}   \quad
	\frac{\mathbf{m}_r^H\tilde{\mathbf{H}}_r\mathbf{m}_r}{\mathbf{m}_r^H\hat{\mathbf{H}}_r\mathbf{m}_r}+\frac{\mathbf{m}_t^H\tilde{\mathbf{H}}_t\mathbf{m}_t}{\mathbf{m}_t^H\hat{\mathbf{H}}_t\mathbf{m}_t} \nonumber\\
	&=\underset{\mathbf{m}_{r}}{\max}   \quad
	\frac{\mathbf{m}_r^H\tilde{\mathbf{H}}_r\mathbf{m}_r}{\mathbf{m}_r^H\hat{\mathbf{H}}_r\mathbf{m}_r}+\underset{\mathbf{m}_{r}}{\max}  \quad \frac{\mathbf{m}_t^H\tilde{\mathbf{H}}_t\mathbf{m}_t}{\mathbf{m}_t^H\hat{\mathbf{H}}_t\mathbf{m}_t}. \label{P1.2}
\end{align}\par
\begin{lemma}\label{lemma1}\it 
	For a Hermitian matrix $\mathbf{A}\in \mathbb{C}^{N\times N}$ and a positive semidefinite matrix $\mathbf{B}\in \mathbb{C}^{N\times N}$, we have
	\begin{align}
		\underset{\left\| \mathbf{\mu}\right\| =p_1}{\arg \max} \quad \frac{\mathbf{\mu}^H\mathbf{A}\mathbf{\mu}}{\mathbf{\mu}^H\mathbf{B}\mathbf{\mu}} = \frac{p_1\mathbf{\xi}_{\max}\left( \mathbf{A}^{-1}\mathbf{B}\right) }{\left\|\mathbf{\xi}_{\max}\left( \mathbf{A}^{-1}\mathbf{B}\right)  \right\| }, \label{D1}
	\end{align}
	where $\mathbf{\xi}_{\max}\left( \mathbf{A}^{-1}\mathbf{B}\right)$ represents the eigenvector corresponding to the maximum eigenvalue of the matrix $\mathbf{A}^{-1}\mathbf {B}$. 
\end{lemma}\par 
\begin{IEEEproof}
	See Appendix \ref{appendixA}.
\end{IEEEproof}\par
Observing \eqref{P1.2}, $\tilde{\mathbf{H}}_r$ and $\tilde{\mathbf{H}}_t$ are Hermitian matrices, and $\hat{\mathbf{H}}_r$ and $\hat{\mathbf{H}}_t$ are positive semidefinite matrices. According to Lemma \ref{lemma1}, the optimal receiving vectors of sensing elements are  $\mathbf{m}_d^*=\sqrt{P_{\rm{s}}}\overline{\mathbf{m}}_d/\left\|\overline{\mathbf{ m}}_d \right\|$, $\forall d$, where $\overline{\mathbf{m}}_d $ is the eigenvector corresponding to the maximum eigenvalue of matrix $\hat{\mathbf{H}}_d^{-1}\tilde{\mathbf{H}}_d$.
\subsection{Transmit Beamforming Optimization}\label{S3B}
Given the sensing filter $\left\{\mathbf{m}_r, \mathbf{m}_t\right\}$ and the MF-RIS reflection/refraction coefficients $\left\{\mathbf{\Theta}_r, \mathbf{\Theta}_t\right\}$, the transmit beamforming optimization problem is formulated as
\begin{subequations}
	\begin{align}
		\underset{\mathbf{w}_k, \mathbf{f}_n}{\max} \quad   
		&\sum_{d\in \mathcal{D}}\gamma_d \label{P2a}	\\
		\mathrm{s.t.}\quad
		&\sum_{k\in\mathcal{K}}\left\| \mathbf{w}_k\right\| ^2 +\sum_{n\in\mathcal{N}} \left\| \mathbf{f}_n\right\| ^2  \le P^{\max}_{\rm{b}}\label{P2b},\\
		&R_{k} \ge R_{\rm{th}}, \forall k\in\mathcal{K}_d,\forall d\in\mathcal{D}\label{P2c},\\
		&	\sum_{d\in\mathcal{D}}\sum_{k\in\mathcal{K}}\left\| \mathbf{\Theta}_d\mathbf{H}\mathbf{w}_k\right\| ^2+\sum_{d\in\mathcal{D}}\sum_{n\in\mathcal{N}}\left\| \mathbf{\Theta}_d\mathbf{H}\mathbf{f}_n\right\| ^2\nonumber\\
		&+\sigma_r^2\sum_{d\in\mathcal{D}}\left\|\mathbf{\Theta}_d \right\| ^2 \le P^{\max}_{\rm{r}}.\label{P2d}
	\end{align}\label{P2}%
\end{subequations}
Problem \eqref{P2} is challenging to solve due to the non-convex objective function \eqref {P2a} and constraint \eqref {P2c}. To transform it into a more tractable form, we reformulate the objective function \eqref{P2a} using the fractional programming method. We define
\begin{align}
	\Delta_d =& \sum_{k\in\mathcal{K}}\left|\mathbf{m}_{d}^H\mathbf{H}_d\mathbf{w}_k\right|^2 + \sum_{n\in\mathcal{N}}\left|\mathbf{m}_{d}^H\mathbf{H}_d\mathbf{f}_n\right|^2\nonumber\\
	 &+ \sigma^2_r\left\| \mathbf{m}_{d}^H\mathbf{G}_d\mathbf{\Theta}_d\right\| ^2 + \sigma^2_s\left\| \mathbf{m}_{d}^H\mathbf{I}_{M_s}\right\| ^2,
\end{align} 
and introduce auxiliary variables $\left\{ \lambda_{d,k}\right\}$ and $\left\{ \eta_{d,k} \right\}$.
\begin{figure*}[t]
	\begin{align}
		\overline{\gamma}_d
		=&\sum_{k\in\mathcal{K}}\left(2\mathrm{Re}\left\{ \lambda_{d,k}^*\mathbf{m}_d^H\mathbf{G}_d\mathbf{\Theta}_d\mathbf{H}\mathbf{w}_k\right\}-\left| \lambda_{d,k}\right|^2\Delta_d \right)
		+\sum_{n\in\mathcal{N}}\left(2\mathrm{Re}\left\{ \eta_{d,n}^*\mathbf{m}_d^H\mathbf{G}_d\mathbf{\Theta}_d\mathbf{H}\mathbf{f}_n\right\}-\left| \eta_{d,n}\right|^2\Delta_d \right) \nonumber\\
		=&\sum_{k\in\mathcal{K}}2\mathrm{Re}\left\{ \lambda_{d,k}^*\mathbf{m}_d^H\mathbf{G}_d\mathbf{\Theta}_d\mathbf{H}\mathbf{w}_k\right\}
		+\sum_{n\in\mathcal{N}}2\mathrm{Re}\left\{ \eta_{d,n}^*\mathbf{m}_d^H\mathbf{G}_d\mathbf{\Theta}_d\mathbf{H}\mathbf{f}_n\right\}
		-\left(\sum_{k\in\mathcal{K}}\left| \lambda_{d,k}\right|^2 +\sum_{n\in\mathcal{N}}\left| \eta_{d,n}\right|^2\right) \Delta_d.\label{SINR^}
	\end{align}
		\hrulefill
\end{figure*}	
Then, the objective function  \eqref{P2a} is transformed to \eqref{SINR^} at the top of this page.\par
By setting 
\begin{align}
	\frac{\partial \overline{\gamma}_d}{\partial \lambda_{d,k}}=\frac{\partial \overline{\gamma}_d}{\partial \eta_{d,n}}=0,\forall d, \forall k, \forall n,
\end{align}
we have 
\begin{align}
	\lambda_{d,k}^*=\frac{\mathbf{m}_d^H\mathbf{G}_d\mathbf{\Theta}_d\mathbf{H}\mathbf{w}_k}{\Delta_d},\quad
	\eta_{d,n}^*=\frac{\mathbf{m}_d^H\mathbf{G}_d\mathbf{\Theta}_d\mathbf{H}\mathbf{f}_n}{\Delta_d}.\label{eta}
\end{align}\par
For the convenience of solving, we define
\begin{align}
	&\mathbf{P}_d=\mathbf{H}_d^H\mathbf{m}_d\mathbf{m}_d^H\mathbf{H}_d,\\
	&\mathbf{Q}=\mathbf{H}^H\mathbf{\Theta}_{t}^H\mathbf{\Theta}_{t}\mathbf{H}+\mathbf{H}^H\mathbf{\Theta}_{r}^H\mathbf{\Theta}_{r}\mathbf{H},\\
	&\mathbf{T}_k=\left(\mathbf{g}_k^H\mathbf{\Theta}_d\mathbf{H}\right)^H\left( \mathbf{g}_k^H\mathbf{\Theta}_d\mathbf{H}\right).
\end{align}%
\begin{figure*}[t]
	\begin{align}
		&\overline{\gamma}_d = 2\sum_{k\in\mathcal{K}}\mathrm{Re}\left\{ \lambda_{d,k}^*\mathbf{m}_d^H\mathbf{G}_d\mathbf{\Theta}_d\mathbf{H}\mathbf{w}_k\right\}
		+2\sum_{n\in\mathcal{N}}\mathrm{Re}\left\{ \eta_{d,n}^*\mathbf{m}_d^H\mathbf{G}_d\mathbf{\Theta}_d\mathbf{H}\mathbf{f}_n\right\}\nonumber\\
		&\quad\quad-\left(\sum_{k\in\mathcal{K}}\left| \lambda_{d,k}\right|^2 +\sum_{n\in\mathcal{N}}\left| \eta_{d,n}\right|^2\right) \left(\sum_{k\in\mathcal{K}}\mathbf{w}_{k}^H\mathbf{P}_d\mathbf{w}_k+\sum_{n\in\mathcal{N}}\mathbf{f}_{n}^H\mathbf{P}_d\mathbf{f}_n+\sigma^2_r\left\| \mathbf{m}_{d}^H\mathbf{G}_d\mathbf{\Theta}_d\right\| ^2+\sigma^2_s\left\| \mathbf{m}_{d}^H\mathbf{I}_{M_s}\right\| ^2 \right).\label{36}\\
		\quad\quad\quad&\mathbf{w}_k^H\mathbf{T}_k\mathbf{w}_k\ge \left( 2^{R_{\rm{th}}}-1\right) \left(\sum_{i\in\mathcal{K},i\neq k}\mathbf{w}_i^H\mathbf{T}_k\mathbf{w}_i+\sum_{n\in\mathcal{N}}\mathbf{f}_n^H\mathbf{T}_k\mathbf{f}_n+ \sigma^2_r\left\| \mathbf{g}_{k}^H\mathbf{\Theta}_d\right\| ^2+\sigma^2_{k}\right), \label{34}\\
		&\sum_{k\in\mathcal{K}}\mathbf{w}_k^H\mathbf{Q}\mathbf{w}_k +\sum_{n\in\mathcal{N}}\mathbf{f}_n^H\mathbf{Q}\mathbf{f}_n+\sigma_r^2\sum_{d\in\mathcal{D}}\left\|\mathbf{\Theta}_d \right\| ^2 \le P^{\max}_{\rm{r}}, \label{35}
	\end{align}
		\hrulefill
\end{figure*}%
Then, the objective function \eqref{SINR^}, constraints \eqref{P2c}, and constraints \eqref{P2d}  can be rewritten as \eqref{36}, \eqref{34}, and \eqref{35} at the top of the this page, respectively.\par
We observe that constraint \eqref {34} is still non-convex, and its lower bound of the left hand side term can be obtained by the SCA method. Specifically, we perform a first-order Taylor approximation on the left hand side term. Then, the non-convex constraint \eqref{P2c} can be replaced by the following equivalent constraint
\begin{align}
	\mathbf{w}_k^H\mathbf{T}_k\mathbf{w}_k \ge 2\mathrm{Re}\left\{	\mathbf{\overline{w}}_k^H\mathbf{T}_k\mathbf{w}_k \right\}-	\mathbf{\overline{w}}_k^H\mathbf{T}_k\mathbf{\overline{w}}_k,
\end{align}
where $\mathbf{\overline{w}}_k$ is the optimal solution in the previous iteration. Finally, problem \eqref{P2} can be reformulated as
\begin{subequations}
	\begin{align}
		\underset{\mathbf{w}_k, \mathbf{f}_n }{\max} \quad   
		&\sum_{d\in \mathcal{D}}	\overline{\gamma}_d \label{P2.1a}\\
		\mathrm{s.t.}\quad
		&\sum_{k\in\mathcal{K}}\left\| \mathbf{w}_k\right\| ^2 +\sum_{n\in\mathcal{N}} \left\| \mathbf{f}_n\right\| ^2  \le P^{\max}_{\rm{b}},\label{P2.1b}\\
		&2\mathrm{Re}\left\{\mathbf{\overline{w}}_k^H\mathbf{T}_k\mathbf{w}_k \right\}-	\mathbf{\overline{w}}_k^H\mathbf{T}_k\mathbf{\overline{w}}_k\ge \nonumber\\
		& \gamma_{\rm{th}} \left(\sum_{i\in\mathcal{K},i\neq k}\mathbf{w}_i^H\mathbf{T}_k\mathbf{w}_i+\sum_{n\in\mathcal{N}}\mathbf{f}_n^H\mathbf{T}_k\mathbf{f}_n+ \sigma^2\right)  \label{P2.1c},\\
		&\sum_{k\in\mathcal{K}}\mathbf{w}_k^H\mathbf{Q}\mathbf{w}_k +\sum_{n\in\mathcal{N}}\mathbf{f}_n^H\mathbf{Q}\mathbf{f}_n+\sigma_r^2\sum_{d\in\mathcal{D}}\left\|\mathbf{\Theta}_d \right\| ^2 \le P^{\max}_{\rm{r}},\label{P2.1d}
	\end{align}\label{P2.2}%
\end{subequations}
where $ \gamma_{\rm{th}}= 2^{{R}_{\rm{th}}}-1$ and $\sigma^2=\sigma^2_r\left\| \mathbf{g}_{k}^H\mathbf{\Theta}_d\right\| ^2+\sigma^2_{k}$. We observe that problem \eqref{P2.2} is a standard SOCP problem, which can be solved by the CVX toolbox \cite{grant2014cvx}. The details of the proposed SOCP-based SCA algorithm for solving problem \eqref{P2} is given in Algorithm \ref{A1}.
\begin{algorithm}[t]
	\caption{ SOCP-based SCA Algorithm for Solving Problem \eqref{P2}}
	\renewcommand{\algorithmicrequire}{\textbf{Input:}}
	\renewcommand{\algorithmicensure}{\textbf{Output:}}
	\begin{algorithmic}[1]
		\REQUIRE Initialized feasible points $\left\{\{\mathbf{w}_k^{(0)}\},\{\mathbf{f}_n^{(0)}\}\right\}$,  the predefined convergence accuracy $\epsilon_1$ and the iteration index $i_1=0$.
		\REPEAT
		\STATE Compute $\lambda_{d,k}^*$ and $\eta_{d,n}^*$ by \eqref{eta}, respectively;
		\STATE Solve problem \eqref{P2.2} via CVX and obtain the solutions $\{\mathbf{w}_k^{(i_1)}\}$ and $\{\mathbf{f}_n^{(i_1)}\}$;
		\STATE Compute $\gamma^{(i_1)}$ by substituting $\{\mathbf{w}_k^{(i_1)}\}$ and $\{\mathbf{f}_n^{(i_1)}\}$ into \eqref{SINR_d};
		\STATE Update $i_1=i_1+1$;
		\UNTIL{$|\gamma^{(i_1)}-\gamma^{(i_1-1)}|\le\epsilon_1$ or the number of iterations reaches maximum;} 
		\ENSURE  The  local optimal transmit beamforming solutions $\{\mathbf{w}_k\}$ and $\{\mathbf{f}_n\}$.
	\end{algorithmic}
	\label{A1}
\end{algorithm}
\subsection{MF-RIS Coefficients Design}
With given transmit beamforming $\left\{ \mathbf{w}_k, \mathbf{f}_n \right\}$ and sensing filter $\left\{\mathbf{m}_r, \mathbf{m}_t\right\}$, we aim to design the MF-RIS reflection/refraction coefficients $\left\{\mathbf{\Theta}_r, \mathbf{\Theta}_t\right\}$. The MF-RIS coefficients design problem can be expressed as
\begin{subequations} 
	\begin{align}
		\underset{\mathbf{\Theta}_r, \mathbf{\Theta}_t}{\max} \quad   
		&\sum_{d\in \mathcal{D}}\gamma_d   \label{P3a}\\
		\mathrm{s.t.}\quad	
		&R_{k} \ge R_{\rm{th}}, \forall k\in\mathcal{K}_d,\forall d\in\mathcal{D}\label{P3b},\\
		&\beta_m^d,\theta_m^d \in \mathcal{R}_{\rm{MF}},\forall m, \forall d,\label{P3c}\\
		&	\sum_{d\in\mathcal{D}}\sum_{k\in\mathcal{K}}\left\| \mathbf{\Theta}_d\mathbf{H}\mathbf{w}_k\right\| ^2+\sum_{d\in\mathcal{D}}\sum_{n\in\mathcal{N}}\left\| \mathbf{\Theta}_d\mathbf{H}\mathbf{f}_n\right\| ^2\nonumber\\
		&+\sigma_r^2\sum_{d\in\mathcal{D}}\left\|\mathbf{\Theta}_d \right\| ^2 \le P^{\max}_{\rm{r}}.\label{P3d}
	\end{align}\label{P3}%
\end{subequations}
We cannot directly solve problem \eqref{P3} since constraints \eqref {P3b} and \eqref {P3d}, and the fractional objective function \eqref {P3a} are non-convex. To address this challenge, a method similar to one in Section \ref{S3B} is adopted to transform the objective function \eqref {P3a} into a more tractable polynomial form. 
\begin{figure*}[t]
	\begin{align}
		\overline{\gamma}_d=&2\mathrm{Re}\left\{\bm{\theta}_d^H\mathbf{t}_d\right\}
		-\left(\sum_{k\in\mathcal{K}}\left| \lambda_{d,k}\right|^2 +\sum_{n\in\mathcal{N}}\left| \eta_{d,n}\right|^2\right) \left(\sum_{k\in\mathcal{K}}\mathbf{w}_{k}^H\mathbf{P}_d\mathbf{w}_k+\sum_{n\in\mathcal{N}}\mathbf{f}_{n}^H\mathbf{P}_d\mathbf{f}_n+\sigma^2_r\bm{\theta}_d^H\mathbf{G}_{m_d}\bm{\theta}_d +\sigma^2_s\left\| \mathbf{m}_{d}^H\mathbf{I}_{M_s}\right\| ^2 \right). \label{45}
	\end{align}
	\hrulefill
\end{figure*}
Specifically, by substituting equation \eqref{eta} into the objective function \eqref{P3a}, \eqref{P3a} is equivalently transformed into \eqref{45} at the top of this page, where 
\begin{algorithm}[t]
	\caption{SOCP-based SCA Algorithm for Solving Problem \eqref{P3}}
	\renewcommand{\algorithmicrequire}{\textbf{Input:}}
	\renewcommand{\algorithmicensure}{\textbf{Output:}}
	\begin{algorithmic}[1]
		\REQUIRE Initialized feasible points $\{\bm{\theta}_r^{(0)}$, $\bm{\theta}_t^{(0)}\}$,  the predefined convergence accuracy $\epsilon_2$ and the iteration index $i_2=0$.
		\REPEAT
		\STATE Compute $\lambda_{d,k}^*$ and $\eta_{d,n}^*$ by \eqref{eta}, respectively;
		\STATE Solve problem \eqref{P3.2} via CVX and obtain the solutions $\bm{\theta}_r^{(i_2)}$ and $\bm{\theta}_t^{(i_2)}$;
		\STATE Compute $\gamma^{(i_2)}$ by substituting $\bm{\theta}_r$ and $\bm{\theta}_t$ into \eqref{SINR_d};
		\STATE Update $i_2=i_2+1$;
		\UNTIL{$|\gamma^{(i_2)}-\gamma^{(i_2-1)}|\le\epsilon_2$ or the number of iterations reaches maximum;} 
		\ENSURE  The  local optimal coefficient solutions $\bm{\theta}_r$ and $\bm{\theta}_t$.
	\end{algorithmic}
	\label{A2}
\end{algorithm}%
\begin{align}
	&	\mathbf{t}_d=\mathrm{diag}\left( \mathbf{m}_d^H\mathbf{G}_d\right)\mathbf{H}\left( \sum_{k\in\mathcal{K}}\lambda_{d,k}^*\mathbf{w}_k
	+  \sum_{n\in\mathcal{N}} \eta_{d,n}^*\mathbf{f}_n\right) ,\\
	&	\mathbf{G}_{m_d}=\mathrm{diag}\left( \mathbf{m}_d^H\mathbf{G}_d\right) \mathrm{diag}\left( \mathbf{G}_d^H\mathbf{m}_d\right). 
\end{align}\par
For the non-convex constraint \eqref{P3b}, we define
\begin{align}
	\mathbf{S}_k=&\mathrm{diag}\left( \mathbf{g}_k^H\right) \mathbf{H}\mathbf{w}_k\mathbf{w}_k^H \mathbf{H}^H\mathrm{diag}\left( \mathbf{g}_k\right), \\
	\overline{\mathbf{S}}_k=&\mathrm{diag}\left( \mathbf{g}_k^H\right) \mathbf{H}\left( \sum_{i\in\mathcal{K},i\neq k}\mathbf{w}_i\mathbf{w}_i^H+\sum_{n\in\mathcal{N}}\mathbf{f}_n\mathbf{f}_n^H\right)  \mathbf{H}^H\nonumber\\
	&\times \mathrm{diag}\left( \mathbf{g}_k\right) +\sigma_r^2\mathrm{diag}\left( \mathbf{g}_k^H\right)\mathrm{diag}\left( \mathbf{g}_k\right). 
\end{align}
Then, the received SINR of the $k$-th user can be re-expressed as
\begin{align}
	\gamma_k=\frac{\bm{\theta}_d^H\mathbf{S}_k\bm{\theta}_d}{\bm{\theta}_d^H	\overline{\mathbf{S}}_k\bm{\theta}_d+\sigma_k^2},	\forall k\in\mathcal{K}_d, \forall d\in\mathcal{D},
\end{align}
and problem \eqref{P3} is equivalent to 
\begin{subequations}
	\begin{align}
		\underset{\bm{\theta}_r, \bm{\theta}_t}{\max} \quad   
		&\sum_{d\in \mathcal{D}}\overline{\gamma}_d   \label{P3.1a} \\
		\mathrm{s.t.}\quad
		&\left| \bm{\theta}_t\right|_m +\left| \bm{\theta}_r\right|_m  \le \beta^{\max},\forall m,\label{P3.1b} \\
		& \bm{\theta}_r^H\mathbf{U}\bm{\theta}_r+\bm{\theta}_t^H\mathbf{U}\bm{\theta}_t
		\le P^{\max}_{\rm{r}}, \label{P3.1c} \\
		&\bm{\theta}_d^H\mathbf{S}_k\bm{\theta}_d \ge \left(2^{R_{\rm{th}}}-1 \right) \left(\bm{\theta}_d^H	\overline{\mathbf{S}}_k\bm{\theta}_d+\sigma_k^2 \right) ,\label{P3.1d} 
	\end{align}\label{P3.1}%
\end{subequations} 
where 
\begin{align}
	\mathbf{U}&=\sum_{k\in\mathcal{K}}\mathrm{diag}\left( \mathbf{H}\mathbf{w}_k\right) \mathrm{diag}\left( \mathbf{w}_k^H\mathbf{H}^H\right)\nonumber\\
	& +\sum_{n\in\mathcal{N}}\mathrm{diag}\left( \mathbf{H}\mathbf{f}_n\right) \mathrm{diag}\left( \mathbf{f}_n^H\mathbf{H}^H\right) +\sigma_r^2\mathbf{I}_M. 
\end{align}
Problem \eqref{P3.1} is still non-convex, as constraint \eqref{P3.1d} is non-convex. In order to transform \eqref{P3.1d} into convex form, we perform a first-order Taylor approximation to its left hand side term. A lower bound of term $\bm{\theta}_d^H\mathbf{S}_k\bm{\theta}_d$ is given by
\begin{align}
	\bm{\theta}_d^H\mathbf{S}_k\bm{\theta}_d  \ge 2\mathrm{Re}\left\{	\overline{\bm{\theta}}_d^H\mathbf{S}_k\bm{\theta}_d \right\}-	\overline{\bm{\theta}}_d^H\mathbf{S}_k\overline{\bm{\theta}}_d,
\end{align}
where  $\overline{\mathbf{\theta}}_d$ is the optimal solution in the previous iteration. Therefore, problem \eqref{P3.1} can be reformulated as
\begin{subequations}
	\begin{align}
		\underset{\bm{\theta}_r, \bm{\theta}_t}{\max} \quad   
		&\sum_{d\in \mathcal{D}}\overline{\gamma}_d \label{P3.3a} \\
		\mathrm{s.t.}\quad
		&\left| \bm{\theta}_t\right|_m +\left| \bm{\theta}_r\right|_m  \le \beta^{\max},\forall m,\label{P3.2b} \\
		& \bm{\theta}_r^H\mathbf{U}\bm{\theta}_r+\bm{\theta}_t^H\mathbf{U}\bm{\theta}_t
		\le P^{\max}_{\rm{r}}, \label{P3.2c} \\
		&2\mathrm{Re}\left\{	\overline{\bm{\theta}}_d^H\mathbf{S}_k\bm{\theta}_d \right\}-	\overline{\bm{\theta}}_d^H\mathbf{S}_k\overline{\bm{\theta}}_d \ge \gamma_{\rm{th}} \left(\bm{\theta}_d^H	\overline{\mathbf{S}}_k\bm{\theta}_d+\sigma_k^2 \right).\label{P3.2d}
	\end{align}\label{P3.2}%
\end{subequations}
We observe that problem \eqref{P3.2} is a standard SOCP problem, which can be solved by the CVX toolbox \cite{grant2014cvx}. The details of the proposed SOCP-based SCA algorithm for solving problem \eqref{P3} is given in Algorithm \ref{A2}.
\begin{algorithm}[t]
	\caption{ AO Algorithm for Solving Problem \eqref{P}}
	\renewcommand{\algorithmicrequire}{\textbf{Input:}}
	\renewcommand{\algorithmicensure}{\textbf{Output:}}
	\begin{algorithmic}[1]
		\REQUIRE Initialized feasible points $\left\{ \{\mathbf{w}_k^{(0)}\}, \{\mathbf{f}_n^{(0)}\}, \bm{\theta}_r^{(0)}, \bm{\theta}_t^{(0)}, \mathbf{m}_r^{(0)}, \mathbf{m}_t^{(0)} \right\}$,  the predefined convergence accuracy $\epsilon_3$ and the iteration index $i_3=0$.
		\REPEAT
		\STATE With given $\left\{\{\mathbf{w}_k^{(i_3)}\}, \{\mathbf{f}_n^{(i_3)}\}, \bm{\theta}_r^{(i_3)}, \bm{\theta}_t^{(i_3)} \right\} $, update $\mathbf{m}_r^{(i_3+1)} =\mathbf{m}_r^{(i_3)}$ and $\mathbf{m}_t^{(i_3+1)}=\mathbf{m}_r^{(i_3)}$ according to Lemma \ref{lemma1};
		\STATE With given $\left\{\bm{\theta}_r^{(i_3)}, \bm{\theta}_t^{(i_3)}, \mathbf{m}_r^{(i_3+1)}, \mathbf{m}_t^{(i_3+1)}\right\}$, update $\{\mathbf{w}_k^{(i_3+1)}\}=\{\mathbf{w}_k^{(i_3)}\}$ and $\{\mathbf{f}_n^{(i_3+1)}\}=\{\mathbf{f}_n^{(i_3)}\}$ by Algorithm \ref{A1};
		\STATE  With given $\left\{ \{\mathbf{w}_k^{(i_3+1)}\}, \{\mathbf{f}_n^{(i_3+1)}\}, \mathbf{m}_r^{(i_3+1)}, \mathbf{m}_t^{(i_3+1)} \right\}$, update $ \bm{\theta}_r^{(i_3+1)}=\bm{\theta}_r^{(i_3)}$ and $ \bm{\theta}_t^{(i_3+1)}=\bm{\theta}_t^{(i_3)}$ by Algorithm \ref{A2};
		\STATE Update $i_3=i_3+1$;
		\UNTIL{$|\gamma^{(i_3)}-\gamma^{(i_3-1)}|\le\epsilon_3$ or the number of iterations reaches maximum;} 
		\ENSURE Local optima solutions $\{\mathbf{w}_k\}$, $\{\mathbf{f}_n\}$, $\bm{\theta}_r$, $\bm{\theta}_t$, $\mathbf{m}_r$, $\mathbf{m}_t$.
	\end{algorithmic}
	\label{A3}
\end{algorithm}
\begin{prop}\label{Proposition}\it 
	The problem \eqref{P0} converges when the proposed AO algorithm is used.
\end{prop}
\begin{IEEEproof}
	See Appendix \ref{appendixB}.
\end{IEEEproof}\par
Based on the above solutions, the details of the overall AO algorithm for solving the original problem \eqref{P} is given in Algorithm \ref{A3}, and its convergence is analyzed by Proposition \ref{Proposition}. In each iteration of Algorithm \ref{A3}, the computational complexity is divided into three parts. Specifically, in Line $2$, a closed-form solution of the optimal sensing filter vector can be obtained by eigenvalue decomposition with the complexity of $\mathcal{O}\left((JM_s)^3 \right) $. The complexity of solving SOCP problems \eqref{P2.2} and \eqref{P3.2} in Lines $3 $ and $4$ is $\mathcal{O}\left((K+J)^4N^4+(4K+4J)N^2\log(1/\epsilon_1)\right) $ and $\left((K+J)^4M^4+(4K+4J)M^2\log(1/\epsilon_2)\right)$, respectively. Therefore, the computational complexity caused by the sensing function at the MF-RIS can be ignored, as it is very low compared to the optimization for MF-RIS coefficients.
\section{Numerical Results}\label{S4}
\begin{table}[t] 
	\centering 
	\renewcommand\arraystretch{1.2}
	\caption{Simulation Parameters}
	\scalebox{1.1}{
		\begin{tabular}{|c|c||c|c|} \hline 
			\bf{Parameter} & \bf{Value} & 	\bf{Parameter} & \bf{Value} \\ \hline\hline
			$K$ & $4$ & $J$ & $4$  \\ \hline	 
			$M$ & $32$ &  $M_s$ & 8  \\ \hline
			$N$ & $8$ &  $P^{\rm{max}}_{\rm{total}}$ &	$45$ $\mathrm{dBm}$ \\ \hline
			$P^{\rm{max}}_r$ &	$15$ $\mathrm{dBm}$&  $P^{\rm{max}}_b$ &	$30$ $\mathrm{dBm}$ \\ \hline
			$R_{\rm{th}}$ & $1$ $\mathrm{bit/s/Hz}$  & $\epsilon_1$  & $10^{-3} $  \\ \hline
			$\epsilon_2$  & $10^{-3} $  & $\epsilon_3$  & $10^{-3} $  \\ \hline
			$\sigma_k^2$ & $-80$ $\mathrm{dBm}$  & $\sigma_r^2$  & $-80$ $\mathrm{dBm}$  \\ \hline
			$\sigma_s^2$  & $-80$ $\mathrm{dBm}$  & $\kappa$ & $10$ $\mathrm{dB}$\\ \hline
			$h_0$ & $-30$ $\mathrm{dBm}$  & $\alpha$ & $-2.8$ \\ \hline
	\end{tabular}}
	\label{T3}
\end{table}
In this section, simulation results are  provided to verify the effectiveness of the proposed MF-RIS-enabled ISAC system. The MF-RIS and the BS are deployed at $ (0, 0, 5)$ m and $(0, 30, 3)$ m, respectively. $K=4$ communication users and $J=4$ sensing targets are randomly distributed in the full-space, and the range of their coordinates is $(-30\, \rm{m} \le x \le 30\, \rm{m}, -30\, \rm{m} \le y \le30\, \rm{m}, z = 0\,\rm{m} )$. Other system parameters are listed in Table \ref{T3}. To evaluate the performance of the proposed algorithm, we compare the proposed algorithm with other three benchmark algorithms, as outlined in Table \ref{T4}, i.e.,
	\begin{itemize}
		\item\textbf{Exhaustive search-based algorithm \cite{9352948}:} We consider the exhaustive search as a performance upper bound for our proposed algorithm, where the MF-RIS coefficients $\Theta_d$ are determined by traversing the feasible set $\mathcal{R}_\mathrm{MF}$. 
		\item\textbf{SDR-based algorithm \cite{9124713}:} The transmit beamforming and MF-RIS coefficients are optimized by the SDR method, where the rank-one constraints $\mathrm{rank}\{\mathbf{R}\}=1\,( \mathbf{R}=\sum_{k=1}^K\mathbf{w}_k\mathbf{w}_k^H+\sum_{n=1}^N\mathbf{f}_n\mathbf{f}_n^H) $ and $\mathrm{rank}\{\hat{
		\mathbf{\Theta}}_d\}=1\,( \hat{
		\mathbf{\Theta}}_d=\bm{\theta}_d \bm{\theta}_d^H) $ are ignored.
		\item\textbf{Random-based algorithm \cite{10225701}:} The phase shifts and amplitude coefficients of  reflection and refraction elements are selected randomly within the feasible set $\mathcal{R}_\mathrm{MF}$.
	\end{itemize}
\begin{figure}[t]
	\centering{}\includegraphics[width=3.5in]{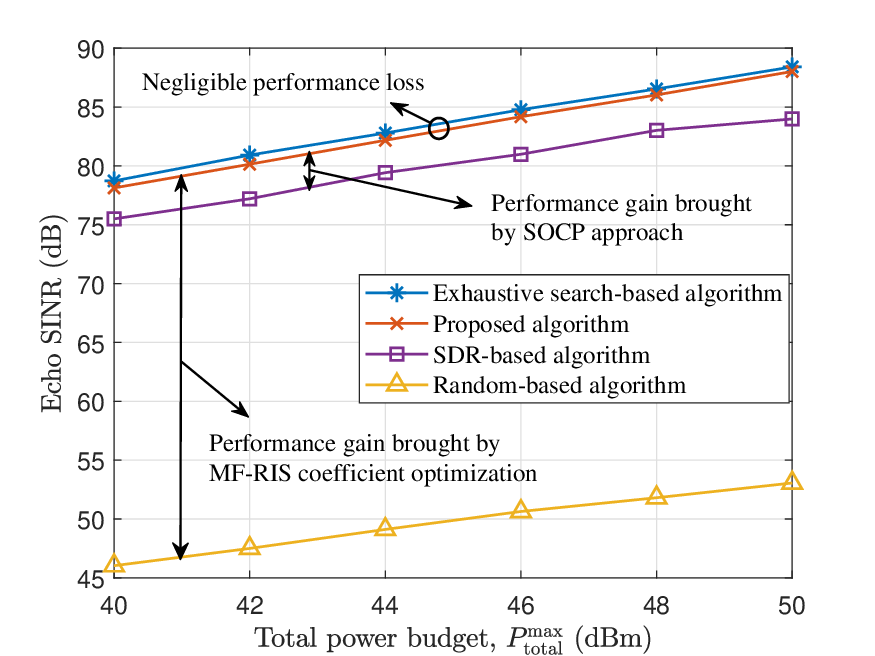}
	\caption{Echo SINR versus $P_{\rm{total}}^{\max}$ under different algorithms.}
	\label{algorithm}
\end{figure}
\begin{figure}[t]
	\centering{}\includegraphics[width=3.5in]{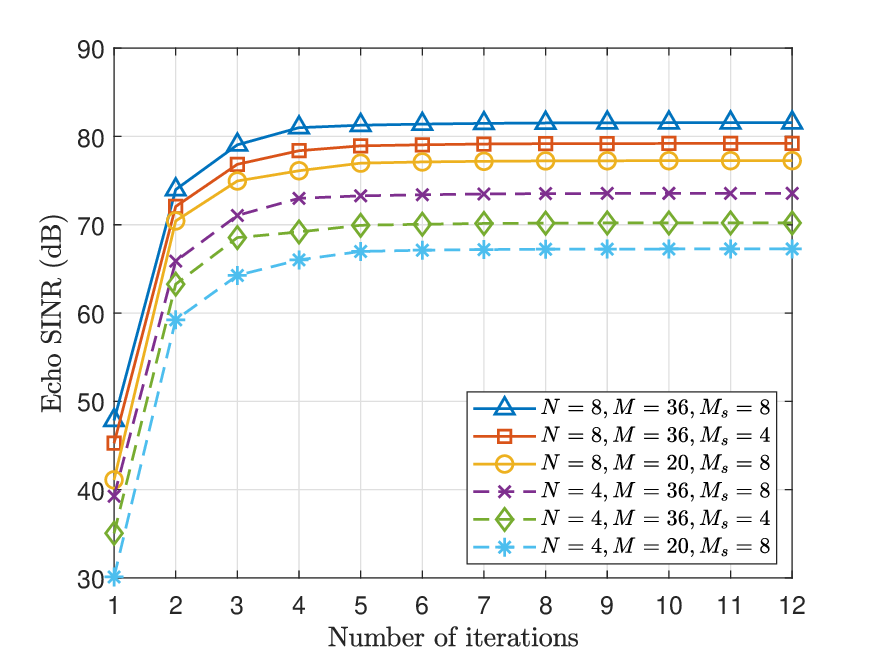}
	\caption{Convergence behavior for the proposed AO algorithm.}
	\label{Convergence}
\end{figure}\par
Fig. \ref{algorithm} presents the echo SINR versus the power budget $P_{\rm{total}}^{\max}$ for different algorithms. The proposed algorithm achieves a higher echo SINR gain than the SDR-based approach. This is because the SDR method often produces high-rank solutions for the relaxed problem, resulting in suboptimal or even infeasible solutions for the original problem. In contrast, the SOCP method employed here effectively converges to a locally optimal rank-one solution. The random-based algorithm performs significantly worse due to the lack of optimization for MF-RIS coefficients. The proposed algorithm delivers performance comparable to the exhaustive search-based method but with much lower computational complexity. Specifically, the complexity of the proposed MF-RIS coefficient algorithm is $\mathcal{O}\left( (4K+4J)M^2\log(1/\epsilon_2)\right)$, while the exhaustive search method, with accuracy $\zeta$, has a complexity of $\mathcal{O}\left( \frac{1}{\zeta^M}\right)$. \par
\begin{table*}[t] 
		\centering 
	\renewcommand\arraystretch{1.2}
	\caption{Benchmark algorithms}
	\scalebox{1.1}{
		\begin{tabular}{|c|c|c|c|} \hline 
			\bf{Algorithm} & \bf{Sensing filter design} & 	\bf{Transmit beamforming optimization} & \bf{MF-RIS coefficients design} \\ \hline
			Proposed algorithm&Lemma \ref{lemma1} & Algorithm \ref{A1}   &Algorithm \ref{A2}    \\ \hline	 
			Exhaustive search-based algorithm&Lemma \ref{lemma1}  &Algorithm \ref{A1}   &Exhaustive search    \\ \hline
			SDR-based algorithm&Lemma \ref{lemma1}  & SDR  &SDR	 \\ \hline
			Random-based algorithm&Lemma \ref{lemma1} 	&  Algorithm \ref{A1}  & Random coefficient\\ \hline
	\end{tabular}}
	\label{T4}
\end{table*}
In Fig. \ref{Convergence}, we present the convergence behavior of the proposed AO algorithm. We observe that the echo SINR received by the MF-RIS elements continues to increase as the number of iterations increases, which verifies the robust convergence of our proposed AO algorithm. By varying the values of $N$, $M$, and $M_s$,  it is evident that the variation in the number of transmit antennas and MF-RIS elements only affects the computational complexity of the proposed algorithm but the convergence is maintained.\par

To demonstrate the benefits brought by the MF-RIS, we consider the proposed scheme for ES, MS, and TS protocols, which are compared with the other three benchmarks. The proposed scheme for three protocols and three benchmark schemes are listed as follows.
\begin{figure*}[htbp]
	\centering
	\begin{minipage}[c]{0.5\textwidth}
		\centering
		\includegraphics[width=3.5in]{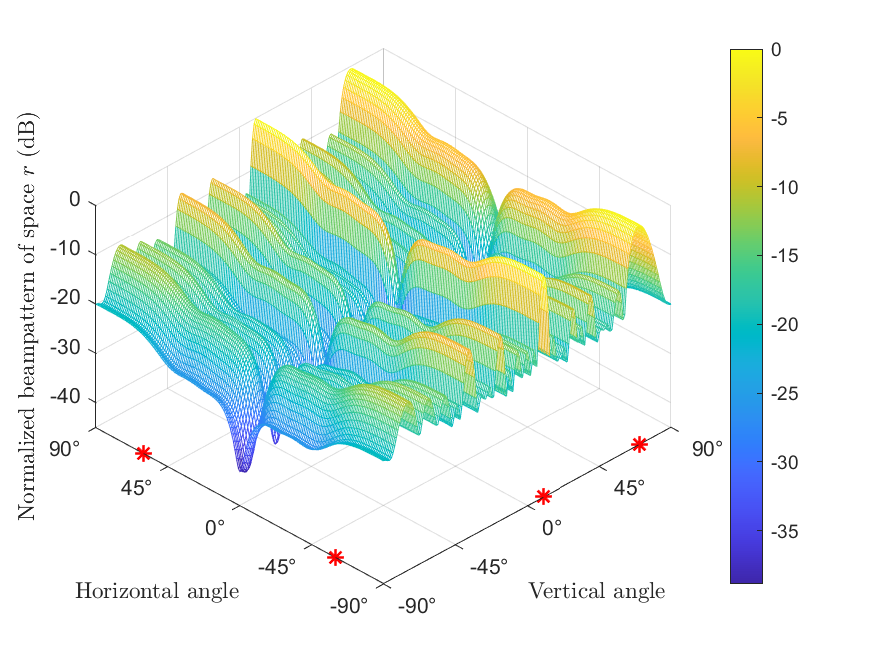}
		\subcaption{3D beampattern  in space $r$.}
		\label{Ba}
	\end{minipage}
	\begin{minipage}[c]{0.49\textwidth}
		\centering
		\includegraphics[width=3.5in]{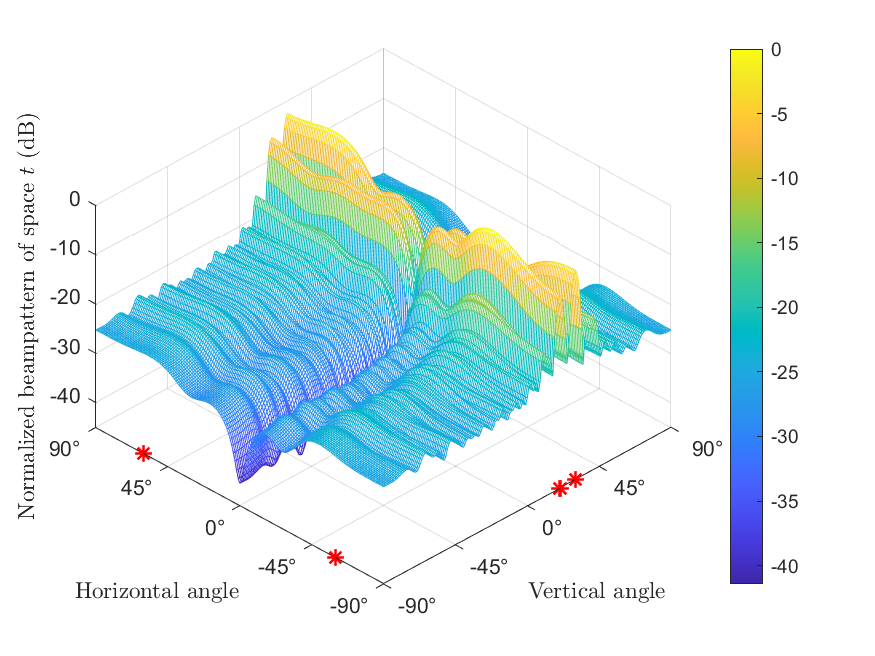}
		\subcaption{3D beampattern  in space $t$.}
		\label{Bb}
	\end{minipage} \\
	\begin{minipage}[c]{0.5\textwidth}
		\centering
		\includegraphics[width=3.5in]{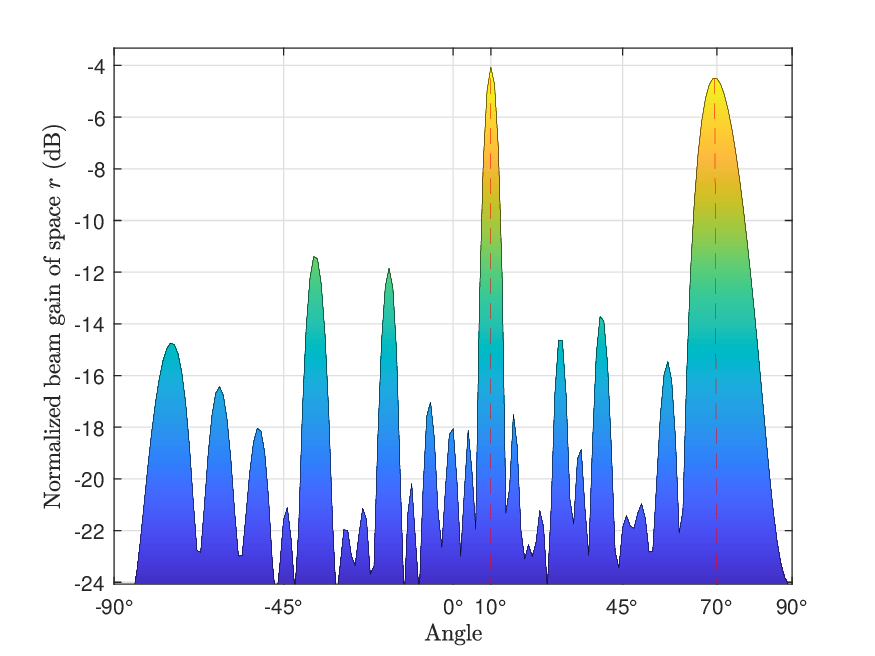}
		\subcaption{2D beampattern  in space $r$.}
		\label{Bc}
	\end{minipage}
	\begin{minipage}[c]{0.49\textwidth}
		\centering
		\includegraphics[width=3.5in]{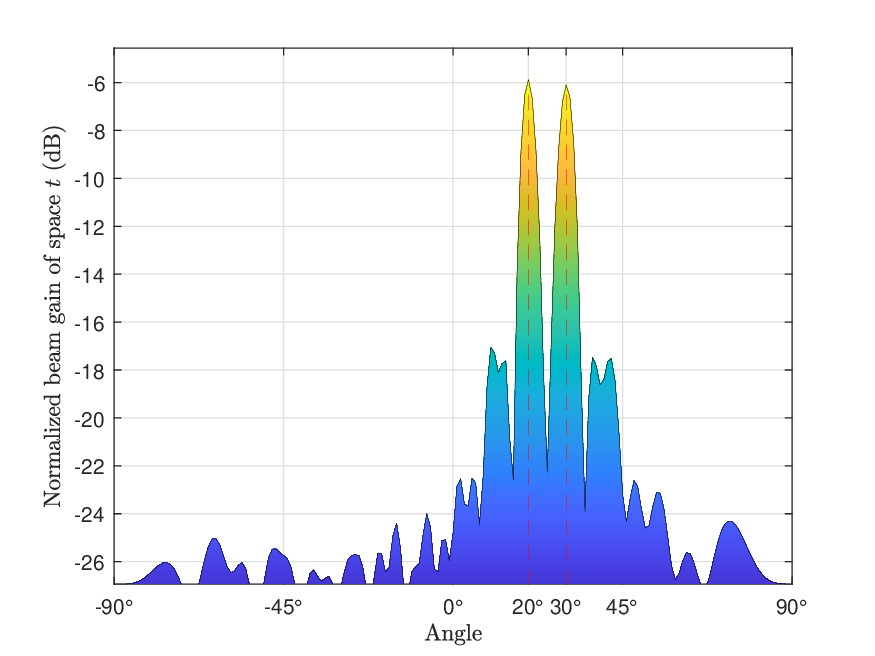}
		\subcaption{2D beampattern  in space $t$.}
		\label{Bd}
	\end{minipage}
	\caption{2D and 3D beampattern in reflection space $r$ and in refraction space $t$.}
	\label{B}
\end{figure*}
\begin{itemize}
	\item \textbf{MF-RIS with ES protocol:} All MF-RIS elements operate in both refraction and reflection modes,  dividing the incident signal into refracted and reflected signals. The corresponding amplitude coefficients satisfy $\beta_m^t+ \beta_m^r \le \beta_{\max}$, $\beta_m^t, \beta_m^r \in \left[ 0,\beta_{\max}\right], \forall m $.
	\item \textbf{MF-RIS with MS protocol:} Each element can only operate in one mode (i.e., refraction or reflection mode). The $M$ MF-RIS elements are divided into two groups: $M_r$ elements operate in the  reflection mode while $M_t$ elements operate in the refraction mode, thus satisfying $M_t+M_r=M$, and $\beta_m^t\beta_m^r =0$, $\beta_m^t, \beta_m^r \in \left[ 0,\beta_{\max}\right], \forall m $. 
	\item \textbf{MF-RIS with TS protocol:} All MF-RIS elements operate in the same mode with period $T$. The time duration that the MF-RIS elements operate in refraction mode is defined as $T_t$ and the time duration that the MF-RIS elements operate in reflection mode is defined as $T_r$, which satisfy $T_t+T_r=T$, $T_t, T_r\in \left[ 0,T\right] $.
	\item \textbf{STAR-RIS \cite{10050406}:} $M$ STAR-RIS elements can refract and reflect the incident signal simultaneously. The amplitude coefficients of STAR-RIS elements satisfy $\beta_m^t, \beta_m^r \in \left[ 0,1\right]$, $\beta_m^t+\beta_m^r=1, \forall m$.
	\item \textbf{Active RIS \cite{10319318}: } $2M$ reflection elements are evenly divided into two equal groups, which serve users and sensing targets in refraction and reflection spaces respectively. The amplitude coefficients of active RIS elements serving refraction space satisfy $\beta_m^r=0, \beta_m^t\in \left[1, \beta_{\max} \right] $ and the amplitude coefficients of active RIS elements serving refraction space satisfy $\beta_m^t=0, \beta_m^r\in \left[1, \beta_{\max} \right] $, where $ m=\{1,2,...,M\}$.
	\item \textbf{Passive RIS \cite{10274660}:} Passive RIS is deployed in the same way as an active RIS, but with different amplitude coefficient settings. The amplitude coefficients of passive RIS elements serving refraction space satisfy $\beta_m^r=0, \beta_m^t=1$ and the amplitude coefficients of passive RIS elements serving refraction space satisfy $\beta_m^t=0, \beta_m^r=1$, where $ m=\{1,2,...,M\}$.
\end{itemize}
\begin{figure}[t]
	\centering{}\includegraphics[width=3.5in]{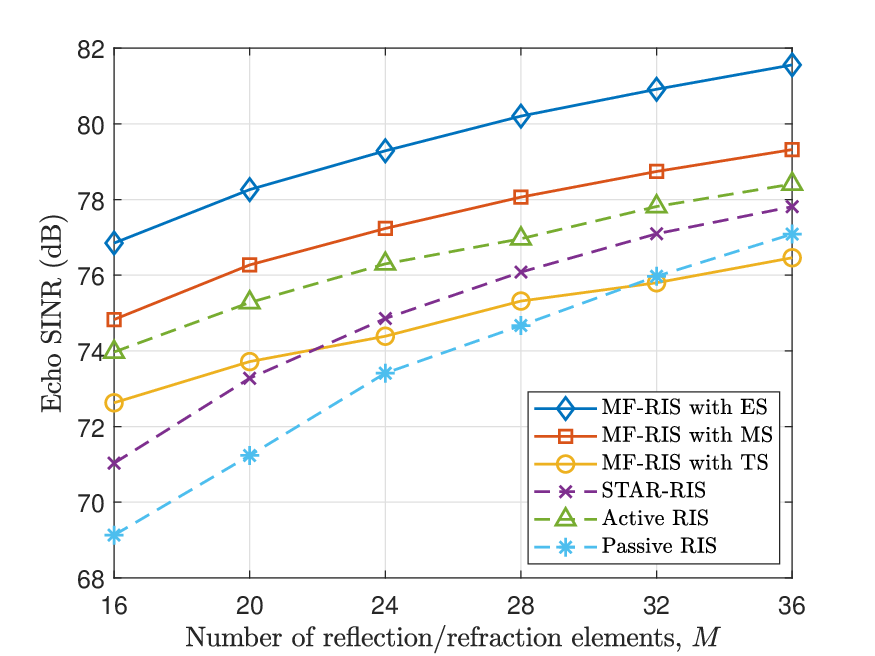}
	\caption{Echo SINR received by sensing elements versus the number of reflection/refraction elements.}
	\label{M}
\end{figure}
\begin{figure}[t]
	\centering{}\includegraphics[width=3.5in]{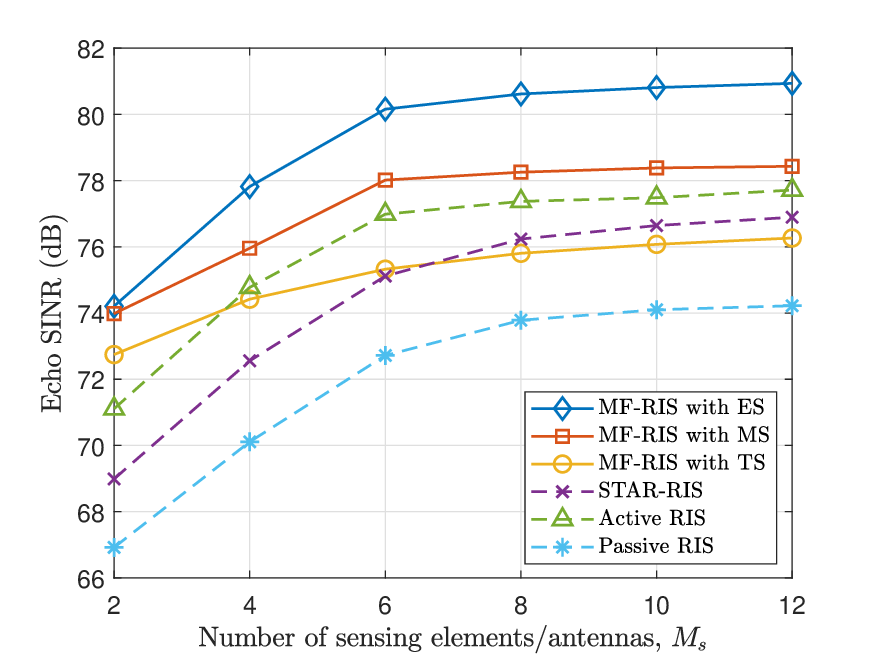}
	\caption{Echo SINR received by sensing elements versus the number of sensing elements/antennas.}
	\label{M_s}
\end{figure}
\begin{figure}[t]
	\centering{}\includegraphics[width=3.5in]{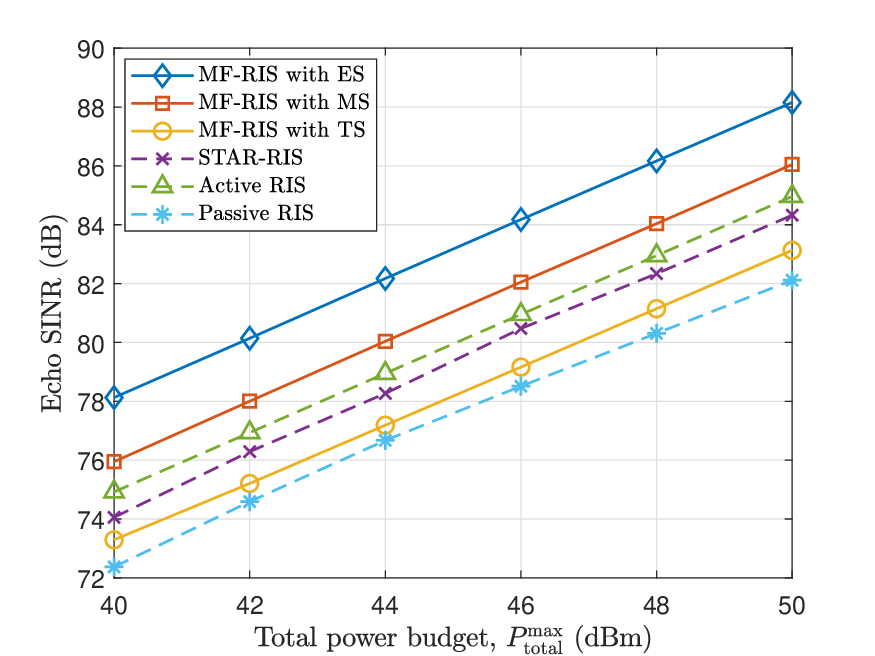}
	\caption{Echo SINR received by sensing elements versus the total power budget.}
	\label{P}
\end{figure}
\begin{figure}[t]
	\centering{}\includegraphics[width=3.5in]{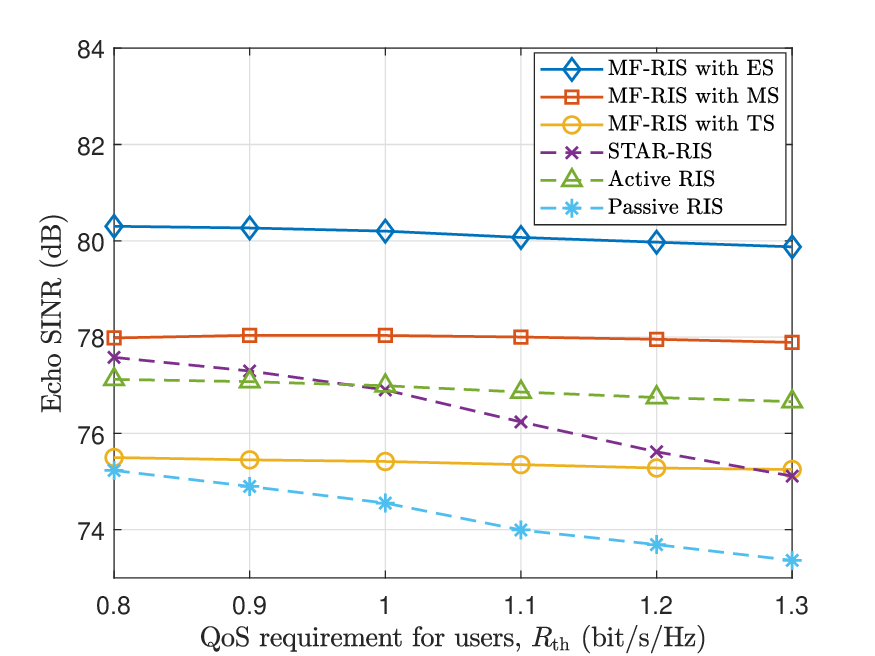}
	\caption{Echo SINR received by sensing elements versus the QoS requirement of user $R_{\rm{th}}$.}
	\label{Rth}
\end{figure}
Note that in MF-RIS-enabled ISAC schemes, the echo is received by the sensing elements, whereas in conventional RIS-assisted ISAC schemes, it is received by the BS.  In addition, for fairness, we define the total power budget as $P^{\max}_{\rm{total}}$, where $P^{\max}_{\rm{b}}+P^{\max}_{\rm{r}}=P^{\max}_{\rm{total}}$ for the MF-RIS schemes and the active RIS scheme, and $P^{\max}_{\rm{b}}=P^{\max}_{\rm{total}}$ for the STAR-RIS scheme and the passive RIS scheme.  \par
Figs. \ref{B} show the  2D and 3D beampattern gain in  reflection space $r$ and refraction space $t$. We assume that horizontal and vertical angles of the four targets are [$60^\circ$,$10^\circ$] and [$-60^\circ$,$70^\circ$] in space $r$, and [$60^\circ$,$20^\circ$] and [$-60^\circ$,$30^\circ$] in space $r$, respectively. The MF-RIS with the ES protocol is adopted. It is clear that the proposed MF-RIS achieves greater beamforming gain towards the directions of targets compared to other directions. This indicates that our proposed MF-RIS scheme can achieve precise target sensing in the full-space while ensuring the quality of communication users.\par
In Fig. \ref{M}, we investigate the impact of the number of reflection/refraction elements on sensing performance. The echo SINR received by sensing elements increases with the number of reflection/refraction elements, owing to the fact that a larger number of such elements offers a greater spatial multiplexing gain for signal transmission. Compared with the TS/MS protocols and three benchmark schemes, the MF-RIS with ES protocol exhibits the best performance. This is due to its ability to provide increased spatial multiplexing, which allows for the elimination of interference through the implementation of reasonable amplitude and phase designs. In addition, passive RIS performs the worst due to the lack of power gain provided by amplification units. Another interesting phenomenon is that as the number of reflection/refraction elements increases, the passive RIS and STAR-RIS outperform the MF-RIS with TS protocol. This indicates that, in the proposed system, time multiplexing provides poorer performance gains compared with spatial multiplexing when the MF-RIS has a large number of reflection/refraction elements.\par
In Fig. \ref{M_s}, we present the impact of the number of sensing elements on sensing performance. It is evident that the system achieves better sensing performance as the number of sensing elements increases. This is because the increased dimensionality of the transmit-receive links brings greater interference cancellation. Both the active RIS and passive RIS schemes exhibit inferior performance compared to MF-RIS schemes with the ES/MS protocol. This is because the beam gain in each half-space is diminished to just a quarter of its original scale when the elements of passive RIS and active RIS are divided evenly into two groups. We observe that the sensing performance of the MF-RIS with the TS protocol is better than the passive RIS but worse than the active RIS. Although spatial multiplexing inherently offers higher beam gain compared to time multiplexing, it cannot compensate for performance loss caused by insufficient power.\par
We plot the relationship between sensing performance and the total power budget for three protocols and two benchmark schemes in Fig. \ref{P}. For all considered protocols and schemes, it is evident that enhancing the total transmit power leads to a more pronounced improvement in sensing performance. This performance gain surpasses the benefits achieved by merely increasing the number of reflection/refraction elements. The MF-RIS with the ES protocol exhibits the best sensing performance compared to other protocols and scheme designs. In contrast, the sensing performance of the MF-RIS with the TS protocol lags behind active RIS and is even comparable with the passive RIS. This further underscores the fact that spatial multiplexing offers a superior sensing performance compared to time multiplexing in the proposed MF-RIS-enabled ISAC system.\par
In Fig. \ref{Rth}, we investigate the impact of the QoS requirement of each user $R_{\rm{th}}$ on sensing performance. Consistent with the observations in Fig. 7, we find that the MF-RIS scheme utilizing the ES protocol achieves the optimal sensing performance, whereas the MF-RIS with the TS protocol is inferior to the active RIS. For the MF-RIS employing the ES, MS, or TS protocols, as well as active RIS, the increasing QoS requirements of users hardly impact the sensing performance. This is attributed to the amplification units equipped in the MF-RIS and active RIS, which provide additional power to compensate for any degradation in communication performance. Conversely, for the passive RIS and STAR-RIS, the sensing SINR declines with rising QoS demands as sensing performance is inevitably sacrificed to fulfill communication requirements due to the lack of power amplification.
\section{Conclusion}\label{S5}
In this paper, we proposed an MF-RIS enabled ISAC system to achieve full-space communication and sensing for multi-user and multi-target scenarios. Our scheme effectively addressed the challenges of limited coverage, multi-hop link fading, and high-cost hardware that are commonly encountered in existing RIS-aided ISAC systems. We formulated an echo SINR maximization problem by jointly optimizing reflection/refraction element coefficients, sensing filter vectors, and the transmit beamforming. This optimization considered the users' QoS requirements, the power budget, and the MF-RIS hardware constraints. An AO algorithm was proposed to efficiently solve this problem by decomposing it into three sub-problems. Simulation results demonstrated that the proposed MF-RIS significantly enhances the sensing performance of ISAC systems while maintaining excellent communication performance compared with existing RIS-aided ISAC systems.
\section*{Acknowledgement}
The authors would like to express their sincere gratitude to the Editor the anonymous Reviewers for their insightful comments and constructive suggestions, which have significantly improved the quality and clarity of this paper. Additionally, we acknowledge that the research proposed by Dr. Wanli Ni in this paper, especially the core concept of MF-RIS, is a continuation of his work during his doctoral studies at Beijing University of Posts and Telecommunications.
\appendices
\section{Proof of Lemma \ref{lemma1}}\label{appendixA}
We construct the left term of \eqref{D1} as the following Rayleigh quotient maximization problem
\begin{subequations}
	\begin{align}
		\underset{\mathbf{\mu}}{ \max} \quad 
		&\frac{\mathbf{\mu}^H\mathbf{A}\mathbf{\mu}}{\mathbf{\mu}^H\mathbf{B}\mathbf{\mu}}\\
		\mathrm{s.t.}\quad
		&\left\| \mathbf{\mu} \right\| =p_1.
	\end{align}\label{Rq}%
\end{subequations}%
If $\mathbf{A}\in \mathbb{C}^{N\times N}$ is a Hermitian matrix, $\mathbf{B}\in \mathbb{C}^{N\times N}$ is a positive semi-definite matrix, we can perform eigenvalue decomposition on $\mathbf{B}$, which can be expressed as
\begin{align}
	\mathbf{B}&=\mathbf{Q} \mathrm{diag}\left( \lambda_1, \lambda_2, ... , \lambda_N\right)\mathbf{Q}^{-1}  \nonumber\\
	&=\left( \mathbf{Q} \mathrm{diag}\left( \sqrt{\lambda_1}, \sqrt{\lambda_2}, ... , \sqrt{\lambda_N}\right)\mathbf{Q}^{H}  \right) ^2,
\end{align}
where $\mathbf{Q}$ is an orthonormal matrix composed of eigenvectors of matrix $\mathbf{A}$, satisfying $\mathbf{Q}^{-1}=\mathbf{Q}^H$. Here, $ \lambda_1, \lambda_2, ... , \lambda_N$ are $N$ eigenvalues of the matrix $\mathbf{A}$. Define $\mathbf{P}= \mathbf{Q} \mathrm{diag}\left( \sqrt{\lambda_1}, \sqrt{\lambda_2}, ... , \sqrt{\lambda_N}\right)\mathbf{Q}^{H}$, then we have $\mathbf{\mu}^H\mathbf{B}\mathbf{\mu}=\left( \mathbf{P}\mathbf{\mu}\right) ^H\mathbf{P}\mathbf{\mu}$. Define $\mathbf{\nu}=\mathbf{P}\mathbf{\mu}$, then \eqref{Rq} can be equivalent to 
\begin{align}
	\underset{\mathbf{\nu}}{ \max} \quad 
	&\mathbf{\nu}^H\mathbf{P}^{-1}\mathbf{A}\mathbf{P}\mathbf{\nu}\\
	\mathrm{s.t.}\quad
	&\left\| \mathbf{\nu} \right\| =p_2.
\end{align}
By constructing the Lagrangian function $L\left(\mathbf{\nu},\lambda \right) = \mathbf{\nu}^H\mathbf{P}^{-1}\mathbf{A}\mathbf{P}^{-1} \mathbf{\nu}-\lambda\left(\left\| \mathbf{\nu} \right\| ^2-p_2^2 \right)  $,  we can obtain the optimal $\mathbf{\nu}$ by solving
\begin{subequations}
\begin{align}
		\frac{\partial L\left(\mathbf{\nu},\lambda \right)}{\partial \mathbf{\nu}}=0,\label{a}\\
		\frac{\partial L\left(\mathbf{\nu},\lambda \right)}{\partial \lambda}=0.\label{b}
\end{align}
\end{subequations}
Then, equations \eqref{a} and \eqref{b} is transformed to equations \eqref{1a} and \eqref{1b}, respectively.
\begin{subequations}
\begin{align}
			&\mathbf{P}^{-1}\mathbf{A}\mathbf{P}^{-1} \mathbf{\nu}=\lambda\mathbf{\nu},\label{1a}\\
		&\left\| \mathbf{\nu} \right\|^2=p_2^2.\label{1b}	
\end{align}
\end{subequations}
Equation \eqref{1a} indicates that the optimal $\mathbf{\nu}^*$ is the eigenvector corresponding to the maximum eigenvalue of the matrix $\mathbf{P}^{-1}\mathbf{A}\mathbf{P}^{-1}$. Equation \eqref{1b} means that the optimal $\mathbf{\nu}$ must satisfy the amplitude constraints. Thus, we can obtain the optimal value $\mathbf{\nu}^*=\frac{p_2\overline{\mathbf{\nu}}}{\left\|\overline{\mathbf{\nu}} \right \| }$, where $\overline{\mathbf{\nu}}$ represents the eigenvector corresponding to the maximum eigenvalue of the matrix $\mathbf{P}^{-1}\mathbf{A}\mathbf{P}^{-1}$. According to the equation $\mathbf{\nu}=\mathbf{P}\mathbf{\mu}$, the optimal solution to problem \eqref{Rq} is $\mathbf{\mu}^*=\frac{p_1\overline{\mathbf{\mu}}}{\left\|\overline{\mathbf{\mu}} \right\| }$, where $\overline{\mathbf{\mu}}$ represents the eigenvector corresponding to the maximum eigenvalue of the matrix $\mathbf {B}^{-1}\mathbf{A}$.
\section{Proof of Proposition \ref{Proposition}}\label{appendixB}
To prove the convergence of the proposed AO algorithm, we denote $ ( \mathbf{m}_r^{(i)}, \mathbf{m}_t^{(i)}, \mathbf{w}_k^{(i)}, \mathbf{f}_n^{(i)}$, $\mathbf{\Theta}_r^{(i)}, \mathbf{\Theta}_t^{(i)} ) $ as the solution of problem \eqref{P} in the $i$-th iteration, where the objective value is defined as $\gamma( \mathbf{m}_r^{(i)}, \mathbf{m}_t^{(i)}, \mathbf{w}_k^{(i)}, \mathbf{f}_n^{(i)}, \mathbf{\Theta}_r^{(i)}, \mathbf{\Theta}_t^{(i)} )$.
Since the local optimal solutions can be obtained at each iteration, we have the following inequality
\begin{align}
	&\gamma\left(\mathbf{m}_r^{(i)}, \mathbf{m}_t^{(i)}, \mathbf{w}_k^{(i)}, \mathbf{f}_n^{(i)},  \mathbf{\Theta}_r^{(i)}, \mathbf{\Theta}_t^{(i)} \right)\nonumber\\
	\overset{(a)}{=}&\gamma\left(\mathbf{m}_r^{(i+1)}, \mathbf{m}_t^{(i+1)}, \mathbf{w}_k^{(i)}, \mathbf{f}_n^{(i)},  \mathbf{\Theta}_r^{(i)}, \mathbf{\Theta}_t^{(i)} \right)\nonumber\\
	\overset{(b)}{\le}&\gamma\left( \mathbf{m}_r^{(i+1)}, \mathbf{m}_t^{(i+1)}, \mathbf{w}_k^{(i+1)}, \mathbf{f}_n^{(i+1)}, \mathbf{\Theta}_r^{(i)}, \mathbf{\Theta}_t^{(i)} \right)\nonumber\\
	\overset{(c)}{\le}&\gamma\left( \mathbf{m}_r^{(i+1)}, \mathbf{m}_t^{(i+1)}, \mathbf{w}_k^{(i+1)}, \mathbf{f}_n^{(i+1)}, \bm{\theta}_r^{(i+1)}, \bm{\theta}_t^{(i+1)} \right)\nonumber\\
	\overset{(d)}{=}&\gamma\left(\mathbf{m}_r^{(i+1)}, \mathbf{m}_t^{(i+1)}, \mathbf{w}_k^{(i+1)}, \mathbf{f}_n^{(i+1)},  \mathbf{\Theta}_r^{(i+1)}, \mathbf{\Theta}_t^{(i+1)} \right), 
\end{align}
where equality (a) holds since the optimal $\mathbf{m}_r^*$ and $\mathbf{m}_t^*$ can be derived by Lemma \ref{lemma1} with given $\{ \mathbf{w}_k^{(i)}, \mathbf{f}_n^{(i)},  \mathbf{\Theta}_r^{(i)}, \mathbf{\Theta}_t^{(i)} \}$, which is proved in Appendix \ref{appendixA}. Inequality (b) holds since the solutions $\mathbf{w}_k^{(i+1)}$ and $\mathbf{f}_n^{(i+1)}$ can be obtained by solving problem \eqref{P2} with given $\{\mathbf{m}_r^{(i+1)}, \mathbf{m}_t^{(i+1)}, \mathbf{\Theta}_r^{(i)}, \mathbf{\Theta}_t^{(i)} \}$, and inequality (c) holds since the solutions $\bm{\theta}_r^{(i+1)}$ and $\bm{\theta}_t^{(i+1)}$ can be obtained by solving problem \eqref{P3} with given $\{ \mathbf{w}_k^{(i+1)}, \mathbf{f}_n^{(i+1)}, \mathbf{m}_r^{(i+1)}, \mathbf{m}_t^{(i+1)} \}$. 
The last equality (d) comes from the fact that $\bm{\theta}_r^{(i+1)}$ and $\bm{\theta}_t^{(i+1)}$ can be equivalently replaced by $\mathbf{\Theta}_r^{(i+1)}$ and $\mathbf{\Theta}_t^{(i+1)}$, respectively. 
Since problem \eqref{P1} has a  closed-form solution, and Algorithms \ref{A1} and \ref{A2}  converge to a local optimum, the objective value of problem \eqref{P0} increases with each iteration. Additionally, the optimum power constraints \eqref{P0b} and \eqref{P0e} ensure that the objective value is bounded above. Therefore, the AO algorithm is guaranteed to converge.

\bibliographystyle{IEEEtran}
\bibliography{Refs}

\end{document}